\documentclass[aps,prr,reprint,longbibliography,superscriptaddress]{revtex4-2}

\usepackage{graphicx}
\usepackage{amsmath}
\usepackage{amssymb}
\usepackage{bm}
\usepackage{subcaption}
\usepackage{hyperref}
\usepackage{xcolor}

\usepackage[normalem]{ulem}

\begin{document}

\title{Spin SYK-dot chains: entanglement bound and resource estimates towards a quantum advantage}

\author{Sayantan Dey}
\email{sayantand84@gmail.com}
\affiliation{Center for Quantum Engineering, Research, and Education, TCG CREST, Kolkata - 700091, India.}

\author{Abel Carreras}
\affiliation{Donostia International Physics Center (DIPC), 20018 Donostia, Euskadi, Spain}

\author{Siddhartha Patra}
\email{siddhartha.patra@tcgcrest.org}
\affiliation{Center for Quantum Engineering, Research, and Education, TCG CREST, Kolkata - 700091, India.}

\begin{abstract}
Ground states of Sachdev--Ye--Kitaev (SYK)-type models are volume-law
entangled, which makes them costly both for tensor networks and for quantum hardware simulations. Thus, it remains unclear where quantum hardware could outperform classical simulation of such models. In this work, we study a spin SYK-dot chain, in which $q$ dense SYK dots of $p$ qubits are joined by single $ZZ$ bonds of strength $K$. The entanglement across any cut is then bounded by $\lfloor p/2\rfloor\ln2$ plus the inter-dot entanglement, and the two-qubit depth of one Trotter step is set by $p$ alone. A resource estimate shows that a quantum advantage for the ground state could appear only for dots of about $15$ to $21$ qubits or more. There, a single Trotter step already requires two-qubit error rates of order $10^{-6}$, beyond the reach of noisy intermediate-scale quantum (NISQ) hardware. We test the assumptions of this estimate with an imaginary-time matrix product state (MPS) that applies each dot as one block. It agrees with exact diagonalisation at $(p,q)=(6,2)$. {At $(5,2)$, a variational circuit with the same dot-plus-link layout comes within $3.6\%$ of the exact energy, and noisy simulations show that its fidelity decays as assumed in the resource estimate.} At $(8,5)$, i.e., $N=40$ qubits, the energy converges at bond dimension $\chi=768$, far below the exact value $2^{20}$, and the inter-dot entanglement stays small and is set mainly by $K$. Out-of-time-order correlators across the links decay on a time scale that shrinks rapidly with $|K|$. Thus, the dot size $p$ and the inter-dot coupling $K$ together set the entanglement of the chain, and with it where a quantum advantage could appear.
\end{abstract}

\maketitle
\raggedbottom

\section{Introduction}
\label{sec:intro}

The Sachdev--Ye--Kitaev (SYK) model, composed of $N$ Majorana fermions with random all-to-all four-body interactions~\cite{SY93,Kitaev15,MS16}, is a solvable model of strongly correlated quantum matter. It is maximally chaotic~\cite{MSS16,SS14}, and is conjectured to possess a holographic dual~\cite{MS16,Chowdhury22}. Further, coupled SYK systems have served as toy models of coupled black holes and traversable wormholes~\cite{MQ18,GJW17}. As these properties appear in strongly entangled states whose exact classical simulation is exponentially costly in $N$, SYK-type models are a natural target for quantum simulation in showing quantum advantage.

On qubits, the Majorana operators of SYK require Jordan-Wigner strings. For this reason, several recent works have turned to Pauli-spin (or ``bosonic'') variants, in which random four-body products of Pauli operators play the role of the four-fermion terms~\cite{Hanada24,Watanabe23,SW24}. SYK dynamics and state preparation have also been studied on noisy quantum processors~\cite{Asaduzzaman24,Araz24,Jha25,Granet25}, including measurements of out-of-time-order correlators (OTOCs) in sparsified bosonic SYK models on IBM hardware~\cite{Gautam25}. Importantly, the authors of Ref.~\cite{Gautam25} stress that any claim of quantum advantage for such models requires a dedicated study of classical simulations, e.g., with tensor networks.

Such studies have repeatedly moved the boundary of a quantum advantage. The random-circuit sampling experiment on Google's Sycamore processor~\cite{Arute19} was followed by tensor-network contractions that solved the same sampling task classically~\cite{PZ22,PCZ22}. Similarly, the kicked-Ising dynamics measured on IBM's 127-qubit Eagle processor~\cite{Kim23} was reproduced with tensor networks adapted to the heavy-hex geometry of the device~\cite{Tindall24,Patra24}, and with Pauli-path methods~\cite{BC24}. More generally, tensor networks whose geometry follows the interaction graph~\cite{Patra24,Patra25} can simulate large systems efficiently as long as their entanglement stays limited. Thus, a claim of quantum advantage is meaningful only against a classical method adapted to the structure of the model.

However, dense SYK-type models are difficult to simulate both with tensor networks and on quantum hardware. SYK eigenstates, including the ground state, obey a volume law of entanglement~\cite{LCB18,HG19}. An MPS with bond dimension $\chi$ can carry at most $\ln\chi$ of entanglement across any cut~\cite{Vidal03,Schollwock11}. Thus, volume-law states require $\chi$ to grow exponentially with $N$, and
dedicated architectures are needed even to represent them~\cite{NoRA24}. A similar obstruction appears on the quantum side. While random circuits readily generate volume-law entanglement, the preparation of a
\emph{specific} volume-law ground state with local gates requires a circuit depth that grows with $N$, as each layer of gates changes the entanglement across a cut by a bounded amount. Further, noise limits the usable depth on current devices~\cite{ZSW20}, and the variational preparation of dense SYK ground states has been found to be difficult~\cite{Gupta26}. Sparsification reduces the number of couplings while retaining much of the SYK phenomenology~\cite{Xu20sparse,GarciaGarcia21,Tezuka23,Orman25}, but it does not by itself remove the volume law. Thus, for these models, the classical cost and the circuit depth grow together with $N$, and a regime in which hardware could outperform classical simulation is difficult to locate.

Ref.~\cite{Gautam25} combines the bosonic form of SYK with random sparsification, and measures OTOCs on an IBM processor. They find that on hardware with nearest-neighbour connectivity, SWAP gates dominate the two-qubit gate count, such that the circuit depth grows rapidly with the system size. Already for 12 system qubits, the measured OTOC approaches zero after the first Trotter step for the denser couplings, and error mitigation by renormalisation becomes unstable. They therefore suggest a sparsification adapted to the hardware.

In this work, we design such a \emph{structured} sparsification. Instead of removing couplings at random, we keep the couplings dense within small groups of qubits, and join these groups by a few structured bonds. We consider $q$ bosonic SYK dots of $p$ qubits each, coupled in a chain. Chains of SYK dots
with random inter-dot couplings have been well studied as models of transport and scrambling~\cite{GQS17,GLQ17,SJB17}. Such chains have also been studied numerically with DMRG, e.g., for a Kondo-lattice chain with an SYK cluster on each site~\cite{BZM18}, and the spreading of entanglement and operators has been studied in Brownian SYK chains, where the couplings fluctuate in time~\cite{NPRT25,MR25}. In contrast, our model is built directly from qubits and has static disorder, and we focus on its ground states. Here, we consider the minimal limit of such chains, in which neighbouring dots are joined by a single, non-random $ZZ$ bond of strength $K$ between their boundary qubits. We refer to this model as a \emph{spin SYK-dot chain}, or simply an SYK-dot chain. Each dot is then volume-law entangled internally, but the entanglement across any cut is bounded by an amount set by $p$ and by the entanglement of the links. We refer to this as an \emph{entanglement budget}.

The SYK-dot chain offers three independent knobs: the dot size $p$ controls the intra-dot entanglement, the coupling $K$ controls the inter-dot entanglement, and the number of dots $q$ sets the system size at a fixed budget. Further, the dots evolve in parallel and the links join only neighbouring dots, such that the circuit depth per Trotter step is set by $p$, and does not grow with $q$. Similarly, an MPS that treats each dot as one block has a cost that grows linearly with $q$ at a fixed bond dimension, and exponentially only with $p$. Thus, the dot size $p$ sets the hardness on both sides, and allows us to locate where a quantum advantage could appear. We note that we use SYK chaos only as a motivation.

We focus on the ground state/s. The entanglement budget holds for any pure state, but its inter-dot part depends on the state. In the ground state, the inter-dot entanglement is static and mainly set by the coupling $K$, while each dot can carry up to $\lfloor p/2 \rfloor \ln 2$ of internal entanglement, so the entanglement across the chain is tunable and bounded. In contrast, in real-time evolution the inter-dot entanglement can grow with time, at a rate bounded by $|K|$. Thus, the ground state is a natural setting in which dense SYK-type models can be studied in a controlled way, both with tensor networks and on quantum hardware.

Several questions remain open. First, can a sparsified SYK model keep dense dots with strong internal entanglement, while its ground state remains tractable both on hardware and with tensor networks? How do the costs on the two sides grow with the dot size $p$ and the number of dots $q$, and in which regime could a quantum device outperform classical simulation? Further, how can a classical simulation exploit the block structure of the chain, such that any advantage is judged against a strong classical baseline? Finally, do the ground states satisfy the assumptions behind these resource estimates, i.e., a bond dimension far below the exact one and a small inter-dot entanglement mainly controlled by $K$? Answering these questions is the main goal of our work.

In Sec.~\ref{sec:model}, we define the SYK-dot chain and its symmetries, and show in Sec.~\ref{sec:budget} that the entanglement of every cut is bounded by $\lfloor p/2\rfloor\ln2$ plus the largest inter-dot entanglement [Eq.~\eqref{eq:entbound}]. In Sec.~\ref{sec:method}, we describe the Trotter circuit used in our resource estimates, a variational circuit with the same dot-plus-link layout at $(p,q)=(5,2)$, and our imaginary-time MPS, which applies the Trotter propagator of each dot as one $p$-qubit block. One step of this MPS has the same dot-plus-link layout as one step of the Trotter circuit. In Sec.~\ref{sec:cost}, we compare the depth and cost of the SYK-dot chain, the sparse SYK model of Ref.~\cite{Gautam25} and dense SYK, and the cost of our block MPS with a conventional gate-by-gate MPS and with an MPS that keeps each dot as one site with higher physical dimension, which serves as the classical reference. We estimate that a quantum advantage for the ground state requires dots of about $15$ to $21$ qubits or more if the bond dimension between dots stays modest. There, even a single Trotter step requires two-qubit error rates of order $10^{-6}$. NISQ hardware can run a single step only for $p\le5$ to $8$ at two-qubit error rates of $10^{-3}$ to $10^{-4}$, where classical simulation is cheap, so this regime lies beyond NISQ hardware. In Sec.~\ref{sec:results}, we test the assumptions of these estimates. We validate the MPS against ED at $(p,q)=(5,2)$ and $(6,2)$, and test the fidelity criterion with noisy simulations of the variational circuit at $(5,2)$. With the MPS at $(6,3)$, $(8,2)$ and $(8,5)$, we then study the bond dimension, the ground-state energy, the entanglement across the link and the OTOCs. We also find that, for even $p$, a boundary qubit stays maximally entangled, and the link only redistributes its entanglement. We conclude in Sec.~\ref{sec:discussion}. The derivation of the budget, the convergence checks, the boundary-qubit entanglement and the OTOC fit are in the appendices.

\section{Model and symmetries}
\label{sec:model}

\subsection{Hamiltonian}

\begin{figure}[t]
\centering
\includegraphics{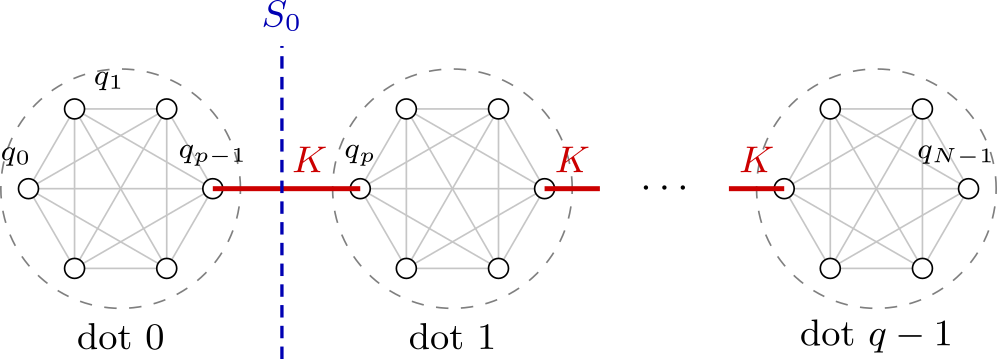}
\caption{Schematic representation of an SYK-dot chain with $q$ dots and $p=6$. Each dashed circle represents a dot containing $p$ qubits with all-to-all four-body couplings (grey lines). Neighbouring dots are joined by a single $ZZ$ bond of strength $K$ (red), for example, between $q_{p-1}$ and $q_p$. The vertical dashed line marks the cut across which we compute the entanglement entropy $S_0$.}
\label{fig:schematic}
\end{figure}

We consider a system of $q$ dots, labelled by $x=0,\dots,q-1$, with $p$ qubits in each dot. Thus, the total number of qubits is $N=pq$. We label the qubits dot by dot as $q_i$, with $i=0,\dots,N-1$, such that the $x$th dot contains $q_{xp},\dots,q_{xp+p-1}$ (Fig.~\ref{fig:schematic}). We write the Hamiltonian as
\begin{equation}
H = \sum_{x=0}^{q-1} H_{\mathrm{SYK}}^{(x)} + H_{\mathrm{int}},
\end{equation}
where
\begin{align}
H_{\mathrm{SYK}}^{(x)} &= \frac{J\sqrt6}{p^{3/2}}
\sum_{\substack{i<j<k<l \\ i,j,k,l\,\in\,\mathrm{dot}\,x}}\;
\sum_{\vec\mu\in\{X,Y\}^4}
J^{(x)}_{ijkl,\vec\mu}\;
\sigma_i^{\mu_1}\sigma_j^{\mu_2}\sigma_k^{\mu_3}\sigma_l^{\mu_4},
\label{eq:Hsyk}\\
H_{\mathrm{int}} &= K \sum_{x=0}^{q-2} Z_{(x+1)p-1} Z_{(x+1)p}.
\label{eq:Hint}
\end{align}
Here, $\sigma^X=X$ and $\sigma^Y=Y$ are Pauli matrices, and the coefficients $J^{(x)}_{ijkl,\vec\mu}\sim\mathcal N(0,1)$ are independent real Gaussian random variables for each Pauli string and each dot. The prefactor follows the usual SYK normalisation~\cite{MS16}, with the number of degrees of freedom replaced by the dot size $p$. Thus, each dot contains $16\binom{p}{4}$ four-qubit Pauli strings. We collect the 16 strings that act on one quadruplet $i<j<k<l$ into $h^{(x)}_{ijkl}$, such that $H^{(x)}_{\mathrm{SYK}}=\sum_{i<j<k<l}h^{(x)}_{ijkl}$. The single-dot Hamiltonian is a restricted version of the SpinXY4 model of Ref.~\cite{Hanada24}, where the four Pauli operators in an interaction term need not act on distinct sites. In contrast, the operators in Eq.~\eqref{eq:Hsyk} act on distinct qubits because $i<j<k<l$.  We impose open boundary conditions and quote all energies per site.

\subsection{Symmetries}
\label{sec:symmetries}
\paragraph*{Dot parities.} Every Pauli string in Eq.~\eqref{eq:Hsyk} contains four $X$ or $Y$ operators within a single dot and therefore commutes with the corresponding dot-parity operator
\begin{equation}
	P_x=\prod_{i\in\mathrm{dot}\,x} Z_i.
\end{equation}
Further, the interaction Hamiltonian in Eq.~\eqref{eq:Hint} consists only of $Z$ operators and commutes with every $P_x$. Thus, $[H,P_x]=0$ for all $x$, and the Hilbert space splits into $2^q$ sectors labelled by the eigenvalues of the dot parities.

\paragraph*{$K\to-K$ after disorder averaging.} Conjugation by $U=\prod_{x=0}^{q-2}X_{(x+1)p}$ reverses the sign of every $ZZ$ bond and of every string with $Y$ on one of these qubits. Thus, $UH(K;\{J\})U^\dagger=H(-K;\{\tilde J\})$, where $\{\tilde J\}$ is as probable as $\{J\}$. Importantly, this symmetry is not exact for a single realisation, e.g., ED at $(p,q)=(5,2)$ gives
$|E(K)-E(-K)|/N$ between $5\times10^{-8}$ and $2\times10^{-5}$ for $K=0.25$ to $2$. We therefore compare $\pm K$ only after disorder averaging.

\paragraph*{Antiunitary symmetry.}The antiunitary $A=\big(\prod_iX_i\big)\mathcal K$, with $\mathcal K$ the complex conjugation, maps $X\to X$, $Y\to Y$ and $Z\to-Z$, and thus commutes 
with $H$ (as the couplings are real). As $AP_xA^{-1}=(-1)^pP_x$, for odd $p$ it maps every parity sector onto the one with all $P_x$ reversed, such that every level is
at least twofold degenerate. The entanglement then depends on which state of this manifold is reached. On the other hand, for even $p$, the ground state of a single dot is generically nondegenerate, and the single-dot version of $A$ forces $\langle Z_a\rangle=0$ in it. 

\subsection{Entanglement budget}
\label{sec:budget}

Let $S_x$ denote the entanglement entropy across the boundary between dots $x$ and $x+1$, with $S_{-1}=S_{q-1}=0$, and let $S_{\mathrm{cut}}$ be that
across a cut after the first $r$ qubits of dot $x$. For a pure state, moving a block of $m$ qubits from one side of a cut to the other changes the entropy
by at most $m\ln2$, by subadditivity and the Araki--Lieb inequality (Appendix~\ref{app:budget}). As the cut differs from the two neighbouring 
dot boundaries by $r$ and $p-r$ qubits,
\begin{equation}
\begin{split}
S_{\mathrm{cut}} &\le \min\{S_{x-1}+r\ln 2,\; S_x+(p-r)\ln 2\}\\
&\le \left\lfloor \tfrac{p}{2}\right\rfloor \ln 2+\max_x S_x
\end{split}
\label{eq:entbound}
\end{equation}
Thus, the entanglement across any cut is bounded by an intra-dot term set by $p$ alone, plus the largest inter-dot entanglement. This holds for any pure
state, with no explicit dependence on $N=pq$. However, $\max_x S_x$ implicitly depends on the full Hamiltonian controlled by $K$. Further, the rate at which 
an interaction generates entanglement is bounded~\cite{VMV13,TB25}.

The same argument, applied to Schmidt ranks, bounds the MPS bond dimension at the cut,
\begin{equation}
\chi_{x,r}\le2^{\lfloor p/2\rfloor}\max\{\chi_{x-1},\chi_x\},
\label{eq:rankbound}
\end{equation}
where $\chi_x$ is the bond dimension at link $x$, in contrast to $\chi_{\mathrm{exact}}=2^{\lfloor N/2\rfloor}$. As the von Neumann entropy does
not strictly control truncation errors, we check the required $\chi$ through convergence (Appendix~\ref{app:convergence}).

\section{Quantum and classical methods}
\label{sec:method}
We prepare the ground state with an MPS on a classical computer and, at $(p,q)=(5,2)$, with a variational circuit that is trained classically and then simulated under gate noise.

\subsection{Quantum circuits}
\label{sec:qmethod}

\paragraph*{Trotter circuit (resource estimates).}
The 16 strings of a quadruplet $h^{(x)}_{ijkl}$ split into two sets of 8 mutually commuting strings, with an even and an odd number of $Y$ operators, respectively.
A Trotter step is the building block of any Trotterised preparation of the ground state on hardware, e.g., by adiabatic evolution. Each weight-four Pauli rotation can be implemented with
a ladder of 6 CNOT gates and single-qubit basis changes. The product of the 8 rotations within one commuting set is exact, such that this implementation
splits each $e^{-i\,\delta t\,h^{(x)}_{ijkl}}$ only into its two commuting halves. Each link gate requires 2 CNOT gates (or one native $ZZ$ rotation). In one Trotter step, all dots act in parallel, followed by one layer of link gates. We use this circuit for the resource estimation in Sec.~\ref{sec:cost}.

\paragraph*{Variational circuit.}
Imaginary-time evolution is not unitary. Quantum imaginary-time evolution~\cite{Motta20} replaces each step by a unitary on a domain that grows with the correlation length, at a cost exponential in its size, here at least a whole dot. A Trotterised preparation needs many steps, each with $962$ two-qubit gates at $(p,q)=(5,2)$ [Eq.~\eqref{eq:n2q}], and a single step already keeps a fidelity of only about $e^{-1}$ at a two-qubit error rate of $10^{-3}$ (Sec.~\ref{sec:cost_map}). We therefore use variational imaginary-time evolution (VarQITE)~\cite{McArdle19}, which gives a shallower circuit. The ansatz has the same dot-plus-link layout as the Trotter circuit. Each of
its $r$ repetitions applies $R_Y$ and $R_Z$ rotations to every qubit and a CNOT
between every pair of qubits within each dot, i.e., an EfficientSU2 layer~\cite{Qiskit24} on
each dot, followed by a trainable $R_{ZZ}$ rotation on each link. A final
rotation layer completes the circuit. Thus, the dots are entangled in parallel,
and the two-qubit depth per repetition is set by $p$, independent of $q$. At
$(p,q)=(5,2)$ with $r=4$, the circuit has $104$ parameters and $88$ CNOT gates,
counting each $R_{ZZ}$ gate as two CNOTs. It is trained entirely on a classical noiseless statevector simulator, with $\Delta\tau=0.2$ and random initial parameters,
until the energy changes by less than $10^{-4}$. We then simulate the trained circuit under gate noise (Sec.~\ref{sec:noisy}).

\subsection{Multi-qubit gates on non-contiguous support}

Each quadruplet term in Eq.~\eqref{eq:Hsyk} acts on four qubits of a dot, which are in general not neighbours in the MPS ordering. A conventional TEBD
implementation would bring them together with $O(p)$ SWAP gates per term~\cite{Paeckel19}. Instead, our simulator, built on the MPS and TEBD
framework of Refs.~\cite{White92,Vidal03,Vidal04,Schollwock11}, applies a gate directly on any set of sites. The gate is contracted with the MPS tensors
spanning these sites, including those on which it acts trivially, and the result is split back through successive singular value decompositions
(SVDs). At each SVD, we keep at most $\chi$ singular values, and drop those below $10^{-10}$ of their norm. We use this to apply the per-dot propagators.

\subsection{Imaginary-time evolution with per-dot block propagators}
\label{sec:imagtime}

We prepare the ground states by imaginary-time evolution,
\[
|\psi(\tau)\rangle \propto e^{-\tau H}|\psi_0\rangle,
\]
using a first-order Trotter decomposition that respects the block structure of the Hamiltonian. For a step $\Delta\tau$, we multiply the exponentials $\exp(-\Delta\tau\,h^{(x)}_{ijkl})$ of all quadruplet terms of dot $x$ exactly into one propagator $M_x(\Delta\tau)$, which acts on the $p$ qubits of that dot. A separate propagator is constructed once for each dot and each step size. One Trotter step is therefore implemented as
\begin{equation}
e^{-\Delta\tau H}\approx
\prod_{x=0}^{q-1} M_x(\Delta\tau)\;
\prod_{x=0}^{q-2}
e^{-\Delta\tau K Z_{(x+1)p-1}Z_{(x+1)p}},
\label{eq:trotter}
\end{equation}
where each $M_x$ is applied as a single $p$-qubit block. The bond dimension $\chi$ thus characterises only the state representation; the propagator itself
is approximated only through the Trotter splitting, both within and between dots. Consequently, the fixed point retains an $O(\Delta\tau)$ Trotter error.
The initial state is chosen as $|\psi_0\rangle=|+\rangle^{\otimes N}$, which has nonzero overlap with every parity sector. Because the Trotter factors preserve these sectors, the state obtained from imaginary time evolution can retain a small weight in sectors above the lowest-energy one when $p$ is even. The convergence with respect to both $\tau$ and $\Delta\tau$ is analysed in Appendix~\ref{app:convergence}.

Real-time evolution, needed for the OTOC (Sec.~\ref{sec:observables}), is implemented using the same block construction, with the replacement $\Delta\tau\rightarrow i\,\delta t$. A propagator over a total time $t$ is consequently decomposed into a sequence
of corresponding Trotter steps.

\subsection{Observables}
\label{sec:observables}

We evaluate three observables from the MPS. The energy is computed term by term,
\[
\langle H\rangle=\sum_S \langle H_S\rangle,
\]
where the sum runs over the individual terms $H_S$ in the Hamiltonian. The quantity $S_0$, introduced in Sec.~\ref{sec:budget}, is extracted from the
Schmidt values $\lambda_i$ across the MPS bond $(p-1,p)$ after bringing the state into mixed canonical form:
\[
S_0=-\sum_i \lambda_i^2\ln \lambda_i^2.
\]

Finally, we compute the out-of-time-order correlator (OTOC) \cite{SS14,Swingle16,XS20} in the ground state $|\psi_{\mathrm{gs}}\rangle$,
\begin{equation}
F(t)=
\langle\psi_{\mathrm{gs}}|
W_t V W_t V
|\psi_{\mathrm{gs}}\rangle,
\label{eq:otoc}
\end{equation}
where
\[
W_t=U_tWU_t^\dagger,
\qquad
U_t=e^{-iHt}.
\]
Operationally, we apply $V$, $U_t^\dagger$, $W$, and $U_t$ successively, and repeat this sequence as required by Eq.~\eqref{eq:otoc}. The resulting overlap
with $|\psi_{\mathrm{gs}}\rangle$ is divided by the norm of the evolved MPS, thereby removing the norm loss caused by truncation. We choose $V=Z_0$ and $W=Z_{N-1}$, so that the two operators act on the opposite ends of the chain. At $K=0$, the dynamics does not couple different dots. Consequently, $W_t$
remains supported on dot $q-1$, including after Trotterisation, while $V$ acts on dot $0$. The two operators therefore commute, and the OTOC satisfies $F(t)=1$ exactly. Our MPS simulations at $K=0$ reproduce this with a deviation of order $10^{-9}$.

\subsection{Relation to other approaches}
Several MPS approaches are available for long-range and multi-qubit terms, e.g., direct MPO application of multi-qubit gates~\cite{PLtensor}, MPO approximations of the propagator of long-range
Hamiltonians~\cite{Zaletel15}, and the time-dependent variational principle~\cite{Haegeman11,Haegeman16}. For ground states, the density-matrix renormalisation group (DMRG)~\cite{White92,Schollwock11} is the standard alternative, and it has been applied to chains of SYK clusters~\cite{BZM18}. For ground-state energies alone, DMRG is more accurate, as it is variational and free of Trotter error. However, our aim is to compare the classical and quantum costs. One Trotter step of our MPS has the same dot-plus-link layout as one step of the Trotter circuit in Sec.~\ref{sec:qmethod}, which we use for cost estimation, such that the classical and quantum costs can be compared side by side (Sec.~\ref{sec:cost}). Further, the same code gives the real-time OTOCs of Sec.~\ref{sec:otoc}.

\section{Resource estimates}
\label{sec:cost}

We now ask where a quantum advantage could arise for the ground state simulation of the SYK-dot chain on NISQ hardware, i.e., without error correction. To address this, we compare the costs of preparing the ground state on quantum hardware versus MPS-based tensor network simulations, as functions of the parameters $p$ and $q$. First, we report the two-qubit gate depth required for a single Trotter step on hardware, and contrast it with the depths for dense and randomly sparsified SYK models. This step is a real-time step, $e^{-i\,\delta t H}$ (Sec.~\ref{sec:qmethod}); as more steps only add gates, its cost is a lower bound on that of a Trotterised preparation. Next, we present the classical computational cost of the block MPS method, and compare it with two alternative MPS approaches. Finally, we combine these hardware and classical costs across the $(p,q)$ parameter plane, incorporating realistic NISQ gate error rates. We do not include the cost of measuring observables. These are operation-count estimates under explicit assumptions, not rigorous bounds on classical or quantum complexity.

\begin{figure}[t]
\centering
\includegraphics[width=\linewidth]{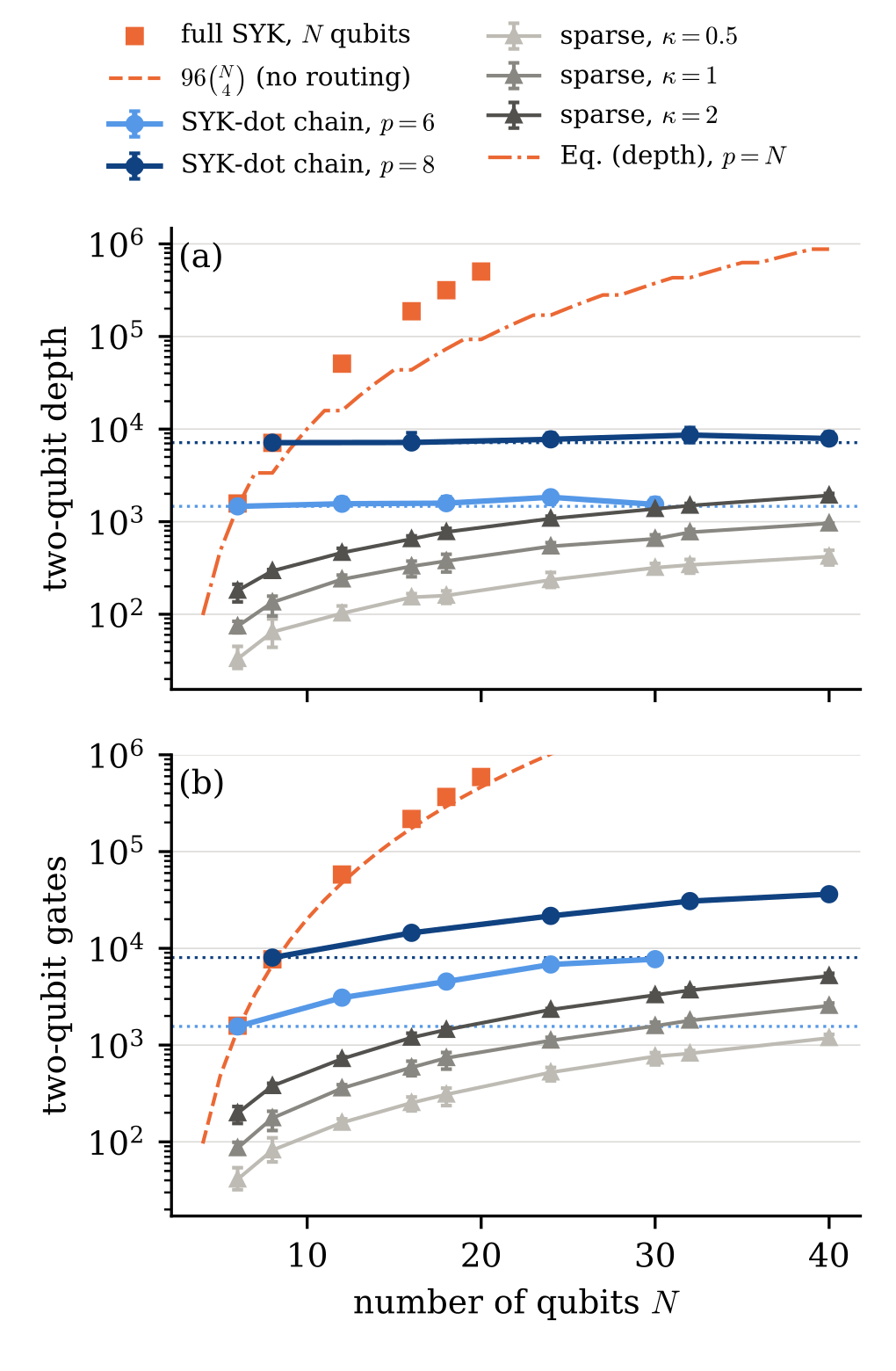}
\caption{Two-qubit cost of one Trotter step after transpilation onto a
156-qubit heavy-hex device (IBM Heron, optimisation level 3): (a) depth and (b)
number of two-qubit gates, against the number of qubits $N$. Circles: SYK-dot
chain for $p=6$ and $8$ with $q=1,\dots,5$, dots placed in order along a path
of the device, mean and range over 10 transpiler seeds. Dotted lines: a single dot, i.e.,
full SYK on $N=p$ qubits. Squares: full SYK on $N$ qubits, mean over 3
transpiler seeds; dash-dotted line:
Eq.~\eqref{eq:depth} with $p=N$ in (a); dashed line: $96\binom N4$ in
(b); both without routing. Triangles: sparse bosonic SYK model of
Ref.~\cite{Gautam25} ($2N$ bosons on $N$ qubits), mean and range over 3
realisations.}
\label{fig:circuitcost}
\end{figure}

\subsection{Hardware: depth per Trotter step}
\label{sec:cost_step}
With the decomposition of Sec.~\ref{sec:qmethod}, the number of two-qubit gates per Trotter step is
\begin{equation}
N_{\mathrm{2q}} = 96\binom p4\,q + 2(q-1),
\label{eq:n2q}
\end{equation}
for all-to-all connectivity within each dot. We do not optimise this count, e.g., by compiling each commuting set jointly, which could reduce it. With linear CNOT ladders and
without gate cancellations, the 96 two-qubit gates of a quadruplet act in sequence. Further, up to $\lfloor p/4\rfloor$ quadruplets on disjoint qubits of one dot can act at the same time, and the dots act on disjoint qubits. Therefore, if the quadruplets can be grouped in this way, the two-qubit depth per Trotter step is
\begin{equation}
D_{\mathrm{2q}} \approx 96\binom p4\Big/\Big\lfloor\frac p4\Big\rfloor + 2,
\label{eq:depth}
\end{equation}
which grows as $16p^3$ at large $p$ and is independent of $q$, for all-to-all connectivity within a dot. For instance, this gives $482$, $1442$ and
$3362$ for $p=5$, $6$ and $8$, respectively. This assumes that neighbouring
dots sit on neighbouring regions of the device, such that SWAP gates are
needed only within a dot. Fig.~\ref{fig:circuitcost} shows the two-qubit depth and gate count of one Trotter step after transpilation onto a 156-qubit heavy-hex device. The depth of the SYK-dot chain stays close to that of a single dot, within $1.5$--$1.8\times10^3$ for $p=6$ and $7.1$--$8.6\times10^3$ for $p=8$, up to $N=30$ and $40$, respectively. In contrast, the depth of the sparse model of Ref.~\cite{Gautam25} grows with $N$, and for sparsity parameter $\kappa=2$  it reaches the value for $p=6$ at $N\approx30$--$35$. For $p=8$, the depth on the device is two to two and a half times that of Eq.~\eqref{eq:depth}, as SWAP gates prevent disjoint quadruplets from acting at the same time. On the other hand, the depth of dense SYK grows rapidly with $N$, from $1.9\times10^5$ at $N=16$ to $5.0\times10^5$ at $N=20$, i.e., about sixty times the depth of the SYK-dot chain with $p=8$ at $N=40$. Thus, among the three models, only the SYK-dot chain keeps the depth per Trotter step fixed as the system grows, although at the sizes studied, the sparse model with $\kappa=2$ is shallower than the chain with $p=8$.

\subsection{Classical: block MPS}
\label{sec:cost_mps}
\label{sec:cost_methods}
We now consider the update of dot $x$. The window of the update is exactly the dot, such that its outer bonds are the link bonds $\chi_{x-1}$ and $\chi_x$ of Eq.~\eqref{eq:rankbound}. The dense $2^p\times2^p$ propagator,  stored once for each step size, is applied to the contracted window at a cost of $\chi_{x-1}\chi_x4^p$. Alternatively, the $\binom p4$ quadruplet gates can be applied one by one, at a cost of $16\binom p4 2^p\chi^2$, which is cheaper only for $p\gtrsim14$. Contracting the window and splitting it back into $p$ tensors through successive SVDs, each truncated to $\chi$, costs $O(2^p\chi^3)$, which exceeds the cost of the propagator for $\chi>2^p${, and a link gate} only $O(\chi^3)$. Therefore, the total cost of the imaginary-time evolution is
\begin{equation}
T_{\mathrm{MPS}} \approx n_{\mathrm{steps}}\,q\Big[\chi^2
\min\Big\{4^p,\;16\binom p4 2^p\Big\}+2^p\chi^3\Big],
\label{eq:tmps}
\end{equation}
where $n_{\mathrm{steps}}$ is the number of Trotter steps{, and the memory is $2^p\chi^2+4^p$ elements, for the window and the propagator.}

Thus, at a fixed $\chi$, both costs grow at most linearly in $q$, while, per Trotter step, they grow polynomially in $p$ on hardware and exponentially in the MPS.

An advantage must be judged against a strong classical method. In the
conventional route~\cite{Paeckel19}, each quadruplet is decomposed into
two-qubit and SWAP gates, each with its own SVD. This gives thousands of
truncations per dot update for $p=6$ and $8$, against $p-1$ for the block
update. In single-core timings of one dot update at $\chi=128$--$256$, the
block update is faster by a factor of $40$ to $60$, with the conventional time
estimated from that of a single gate. The number of gates grows only
polynomially with $p$, while the dense propagator grows as $4^{p}$ (for
$p\lesssim 14$), so this gain decreases rapidly with $p$.

Further, the inner bonds of a middle dot in a volume-law state can require up
to $\chi \approx 2^{\lfloor p/2\rfloor}\,\chi_{\mathrm{link}}$
[Eq.~\eqref{eq:rankbound}], where the link bond dimension
$\chi_{\mathrm{link}}$ is set by the entanglement across the links. If they
reach this bound, the SVD term of Eq.~\eqref{eq:tmps} grows as
$2^{5p/2}\,\chi_{\mathrm{link}}^{3}$ per dot and step. Our converged bond
dimensions are consistent with this bound (Sec.~\ref{sec:bond}).

However, the block MPS is not the cheapest classical method. Its bond dimension can grow as $2^{\lfloor p/2\rfloor}$ only
because each dot is split into $p$ sites. In the same imaginary-time
evolution, each dot can instead be kept as one site of dimension $2^p$. Only
the link bonds then remain, and the quadruplet gates act directly on the site
tensor. The cost per dot and step then grows as
$\binom p4 2^p\chi_{\mathrm{link}}^2$, i.e., as $2^p$ rather than as
$2^{5p/2}$ or faster, and the memory as $2^p\chi_{\mathrm{link}}^2$. We use this variant, which we have not run and whose
cost follows from the operation count, as the classical reference below, with
the same $\chi_{\mathrm{link}}$ as the block MPS. Fewer iterations, or truncating each dot to its lowest
eigenstates, could reduce the cost further.

For comparison, we also give the cost of the other two methods used in
Sec.~\ref{sec:results}.

\paragraph*{Exact diagonalisation.} Even when restricted to one parity sector,
the Hilbert space has dimension $2^{q(p-1)}$, and each Lanczos iteration costs
$O\big((16\binom p4q+q-1)\,2^{q(p-1)}\big)$. The sector dimension is $2^{8}$
and $2^{10}$ for $(p,q)=(5,2)$ and $(6,2)$, respectively, but $2^{35}$ for
$(p,q)=(8,5)$, i.e., about $0.5$\,TB per vector.

\paragraph*{VarQITE.} The ansatz of Sec.~\ref{sec:qmethod} acts on all $N$
qubits, with $n_\theta=2N(r+1)${${}+(q-1)r$} parameters for $r$ repetitions, e.g.,
$n_\theta=${$104$} for $r=${$4$} at $N=10$.

On a statevector simulator, each step of the
McLachlan equations~\cite{McArdle19} requires the $n_\theta\times n_\theta$
quantum geometric tensor, the energy gradient and a linear solve, at a cost of
$O\big((n_\theta^2+M_H)\,2^N+n_\theta^3\big)$, where $M_H$ is the number of
Pauli strings in $H$. Thus, the training cost grows exponentially with $N=pq$,
and not only with $p$. On the other hand, training on hardware would require
each of the $n_\theta(n_\theta+1)/2$ elements of the geometric tensor to be
estimated from a separate circuit, i.e., about $5\times10^3$ circuits per step for
$n_\theta=${$104$}, each with many shots.

\paragraph*{MPS.} 
Finally, at $(p,q)=(8,5)$, the MPS [Eq.~\eqref{eq:tmps}] is the cheapest of the three
methods, as the statevector training of VarQITE would also require $2^{40}$
amplitudes.

\subsection{Where could a quantum advantage appear?}
\label{sec:cost_map}

\begin{figure}[t]
\centering
\includegraphics[width=\linewidth]{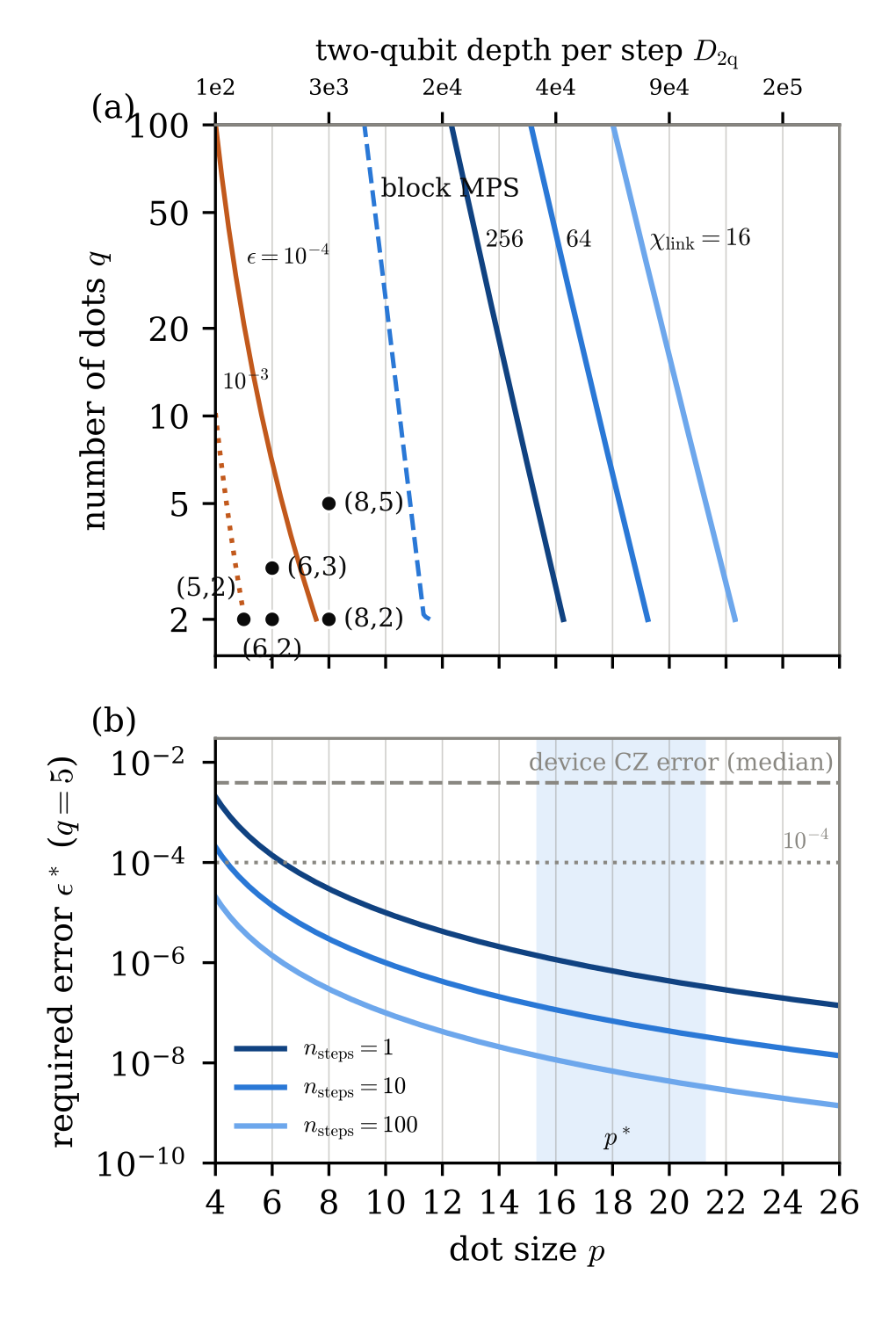}
\caption{Where a quantum advantage could appear. (a) Classical budget
(blue; solid: MPS with one site per dot, dashed: block MPS) and NISQ reach of one Trotter step (orange) in the $(p,q)$ plane;
points: systems studied in this work. (b) Two-qubit error rate $\epsilon^*$
required on hardware, for $q=5$; shaded: $p^*$. }
\label{fig:costmap}
\end{figure}

Figure~\ref{fig:costmap} combines these estimates. As a classical
budget, we take $10^{18}$ operations, i.e., about a day at $10^{13}$
operations per second, and $1$\,TB of memory. Larger machines move the
contours to larger $p$. For the MPS with one site per
dot, this budget gives a contour in the $(p,q)$ plane for each
$\chi_{\mathrm{link}}$, along which the largest $p$ decreases only
logarithmically with $q$. Systems to the left of a contour lie within the
budget. At $q=5$, the contours for $\chi_{\mathrm{link}}=16$ to $256$ lie at
$p^*\approx15$--$21$, far beyond those of the block MPS. Each contour assumes
that its $\chi_{\mathrm{link}}$ suffices. As $(8,5)$ at $|K|=2$ already
requires $\chi_{\mathrm{link}}\ge48$ (Sec.~\ref{sec:bond}), the contour
for $16$ is an optimistic limit for the classical side at this coupling, and
$p^*$ is smaller if $\chi_{\mathrm{link}}$ grows with $p$. Thus, the dot size
$p$, together with the link entanglement set by $K$, decides whether the
ground state lies within the classical budget, while the number of dots
matters much less, as long as $\chi_{\mathrm{link}}$ does not grow with $q$.
Our single realisation at $p=8$ cannot decide this
(Sec.~\ref{sec:bond}), and we have estimates of $\chi_{\mathrm{link}}$
only for $p\le8$.

Beyond the contours, the classical cost of the methods considered here
grows exponentially with $p$, while the two-qubit depth per Trotter step [top
axis of Fig.~\ref{fig:costmap}(a)] grows only as $p^3$, independent of $q$,
for all-to-all connectivity within a dot. A
quantum advantage could thus appear for $p\gtrsim p^*$, provided that the
number of Trotter steps needed to prepare the ground state, e.g.,
adiabatically, does not grow exponentially with $p$. This number is set by
the gap of the dots, which we have not studied.

However, if each two-qubit gate fails independently with probability
$\epsilon$, $n_{\mathrm{steps}}$ Trotter steps keep a fidelity of about
$\exp(-\epsilon\,n_{\mathrm{steps}}N_{\mathrm{2q}})$, which requires
$\epsilon\lesssim\epsilon^*=1/(n_{\mathrm{steps}}N_{\mathrm{2q}})$. A single
step meets this condition only below the orange lines in
Fig.~\ref{fig:costmap}(a), i.e., for $p\le5$ at $\epsilon=10^{-3}$, close to
the best current devices, and for $p\le8$ at $10^{-4}$, with few dots. As
these lines use the gate count of Eq.~\eqref{eq:n2q}, routing on a device
with limited connectivity lowers this reach further
(Fig.~\ref{fig:circuitcost}). Error mitigation can extend this reach, but for
probabilistic error cancellation, the number of samples grows exponentially
with the total error $\epsilon\,n_{\mathrm{steps}}N_{\mathrm{2q}}$, such that
it becomes impractical for a total error much larger than
one~\cite{Temme17,vdBerg23}. At $p\approx p^*$, $\epsilon^*$ is of order
$10^{-6}$ or below already for a single step, i.e., about two orders of
magnitude below $10^{-4}$, and it decreases further with the number of steps
[Fig.~\ref{fig:costmap}(b) and Table~\ref{tab:resources}].

Thus, within these estimates, a quantum advantage for the ground state
requires dots of at least $p^*$ qubits, i.e., about $15$
to $21$ qubits if $\chi_{\mathrm{link}}$ lies between $16$ and $256$, beyond the reach of NISQ hardware, while the systems within its
reach are classically cheap.

\begin{table}[t]

\caption{Resources for one Trotter step of an SYK-dot chain with $q=5$
dots. Hardware: two-qubit depth $D_{\mathrm{2q}}$ [Eq.~\eqref{eq:depth},
all-to-all connectivity within a dot], number of two-qubit gates
$N_{\mathrm{2q}}$ [Eq.~\eqref{eq:n2q}], and the two-qubit error rate
$\epsilon^*=1/N_{\mathrm{2q}}$ required for a fidelity of about $1/e$ after a
single step. Classical: cost
of the ground state in operations, for $\chi_{\mathrm{link}}=64$, with the
block MPS [$T_{\mathrm{block}}$, Eq.~\eqref{eq:tmps}] and with one site per
dot ($T_{\mathrm{dot}}$), both with $n_{\mathrm{steps}}=3000$.}
\label{tab:resources}
\begin{ruledtabular}
\begin{tabular}{ccccccc}
$p$ & $N$ & $D_{\mathrm{2q}}$ & $N_{\mathrm{2q}}$ & $\epsilon^*$ & $T_{\mathrm{block}}$ & $T_{\mathrm{dot}}$ \\
\hline
$6$  & $30$  & $1442$   & $7.2\times10^3$ & $1.4\times10^{-4}$ & $1.4\times10^{14}$ & $2.0\times10^{12}$ \\
$8$  & $40$  & $3362$   & $3.4\times10^4$ & $3.0\times10^{-5}$ & $5.2\times10^{15}$ & $2.2\times10^{13}$ \\
$10$ & $50$  & $10082$  & $1.0\times10^5$ & $9.9\times10^{-6}$ & $2.0\times10^{17}$ & $2.3\times10^{14}$ \\
$12$ & $60$  & $15842$  & $2.4\times10^5$ & $4.2\times10^{-6}$ & $8.4\times10^{18}$ & $2.1\times10^{15}$ \\
$16$ & $80$  & $43682$  & $8.7\times10^5$ & $1.1\times10^{-6}$ & $1.2\times10^{22}$ & $1.2\times10^{17}$ \\
$20$ & $100$ & $93026$  & $2.3\times10^6$ & $4.3\times10^{-7}$ & $9.7\times10^{24}$ & $5.0\times10^{18}$ \\
$24$ & $120$ & $170018$ & $5.1\times10^6$ & $2.0\times10^{-7}$ & $7.5\times10^{27}$ & $1.8\times10^{20}$ \\
\end{tabular}
\end{ruledtabular}
\end{table}

\section{Results}
\label{sec:results}

We now test the assumptions behind the resource estimates of
Sec.~\ref{sec:cost}. First, the block MPS must be exact where it should be.
Second, the required bond dimension must stay within
$2^{\lfloor p/2\rfloor}\chi_{\mathrm{link}}$, with a modest $\chi_{\mathrm{link}}$.
Third, the inter-dot entanglement must stay small and be set by $K$. Finally,
the spreading of operators shows how fast real-time  dynamics leave this controlled regime. We test the first assumption at $(p,q)=(5,2)$ and $(6,2)$, where exact results
exist. We test the other three at $(6,3)$, $(8,2)$ and $(8,5)$, one quantity at
a time. The OTOC is computed only at $(6,3)$ and $(8,2)$.

\subsection{Validation}
\label{sec:validation}

We first test whether the block MPS is exact where it should be, and
how a variational circuit behaves under gate noise. We  use the two
smallest system sizes, where exact results are available.

\subsubsection{$(p,q)=(5,2)$: variational circuit under gate noise}
\label{sec:noisy}

\begin{figure}[t]
\centering
\includegraphics[width=\linewidth]{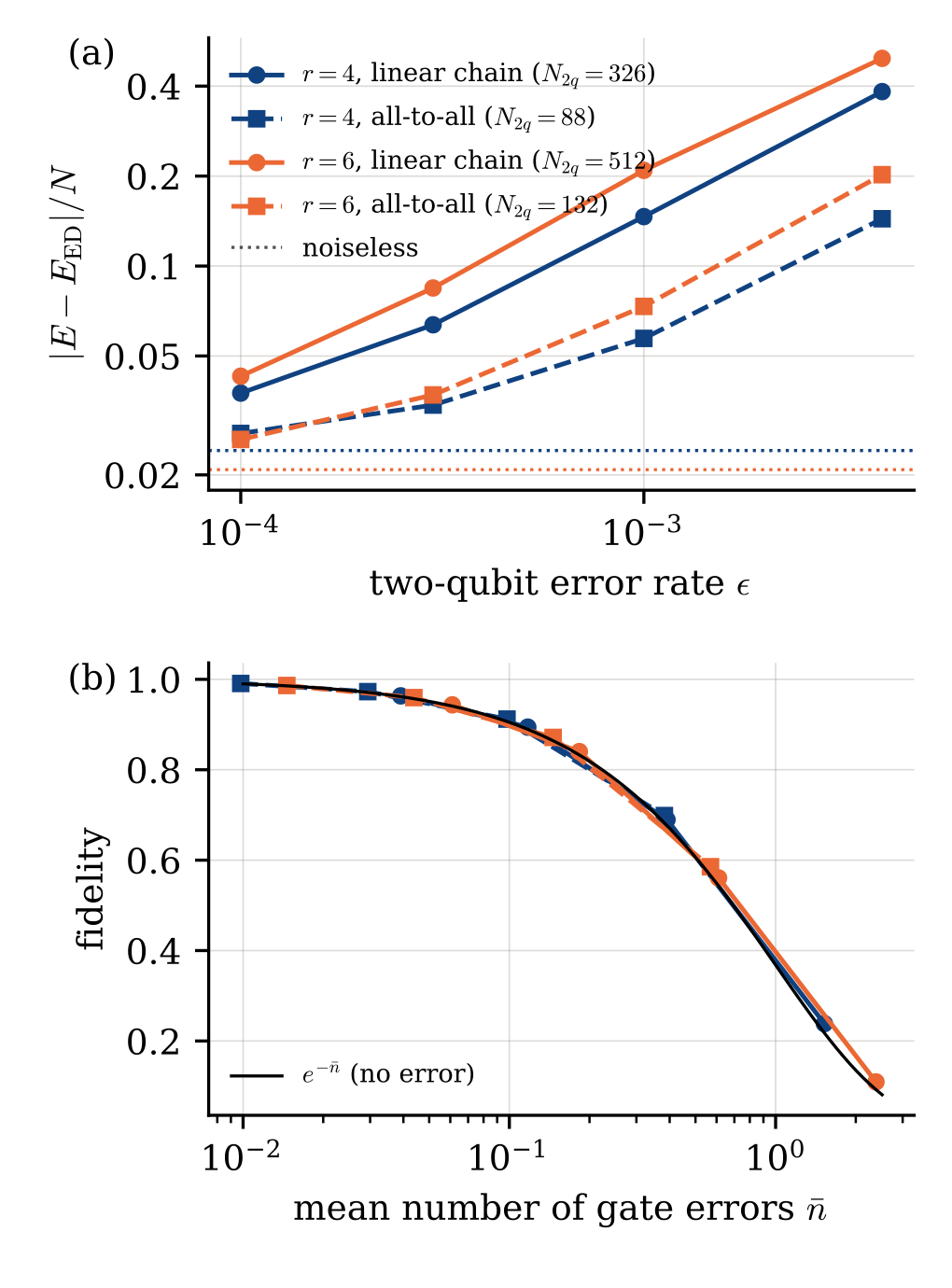}
\caption{Dot-plus-link variational circuit at $(p,q)=(5,2)$, $K=2$,
under depolarising gate noise, for $r=4$ and $6$ repetitions, compiled for
all-to-all connectivity (dashed) and onto a linear chain of a heavy-hex
device (solid). (a) Deviation of the energy per site from ED against the
two-qubit error rate $\epsilon$; dotted: noiseless. (b) Fidelity with the noiseless
state against the mean number of gate errors in one run, $\bar n$; black
line: $e^{-\bar n}$, the probability of no error.}
\label{fig:noisysim}
\end{figure}

For a single disorder realisation at $(p,q)=(5,2)$ and $K=2$, ED gives
$E/N=-0.66800$, and the block MPS agrees to within $10^{-5}$. We use this
system to test the variational circuit of Sec.~\ref{sec:qmethod} and the
fidelity criterion of Sec.~\ref{sec:cost_map}. Trained without noise, the
circuit with $r=4$ reaches $E/N=-0.64387$, i.e., $3.6\%$ above ED, compared
with $20\%$ for $r=2$ and $3.1\%$ for $r=6$. Thus, a few repetitions of the
dot-plus-link layout suffice at this size, consistent with the small
inter-dot entanglement of the ground state (Sec.~\ref{sec:ee}).

We then simulate the trained circuit with the density-matrix simulator
of Qiskit Aer~\cite{Qiskit24}. After every two-qubit gate, a random non-identity two-qubit
Pauli error occurs with probability $\epsilon$, and after every single-qubit
gate, a single-qubit Pauli error with probability $\epsilon_1=\epsilon/10$.
We compile the circuit either for all-to-all connectivity, with $88$ CNOT
gates for $r=4$, or onto a linear chain of ten qubits of the heavy-hex device
of Fig.~\ref{fig:circuitcost}, where routing raises the number of CZ gates to
$326$. On the linear chain, the energy lies $5.6\%$, $9.5\%$ and $22\%$ above
ED for $\epsilon=10^{-4}$, $3\times10^{-4}$ and $10^{-3}$, respectively, and
at $\epsilon=3.9\times10^{-3}$, the median CZ error of this device model,
only $44\%$ of the noiseless energy remains [Fig.~\ref{fig:noisysim}(a)].
Further, the circuit with $r=6$ is more accurate without noise, but has more
two-qubit gates ($132$ instead of $88$ for all-to-all connectivity), and is
thus less accurate at every $\epsilon$ studied. The number of repetitions must
therefore be chosen together with the error rate.

Figure~\ref{fig:noisysim}(b) shows the fidelity
$F=\langle\psi_\theta|\rho_\epsilon|\psi_\theta\rangle$ between the noisy
output $\rho_\epsilon$ of the trained circuit and its noiseless output
$|\psi_\theta\rangle$, against
$\bar n=\epsilon N_{\mathrm{2q}}+\epsilon_1N_{\mathrm{1q}}$, the mean number of
gate errors in one run, where $N_{\mathrm{2q}}$ and $N_{\mathrm{1q}}$ are the
numbers of two- and single-qubit gates. Thus, $F$ measures only the effect of
the noise, not the error of the ansatz itself. The black line, $e^{-\bar n}$,
is the probability of a run without errors. For both circuits and both
layouts, $F$ lies on or slightly above it, by at most $10\%$ for $F>0.2$.
This supports the criterion $\epsilon N_{\mathrm{2q}}\lesssim1$ used in
Fig.~\ref{fig:costmap}. As these simulations include only gate errors, they
give an optimistic estimate for a device.

\subsubsection{$(p,q)=(6,2)$: exact diagonalisation versus MPS}
\label{sec:ed12}

\begin{figure*}[t]
\centering
\begin{subfigure}[t]{0.49\textwidth}
\centering
\includegraphics[width=\linewidth]{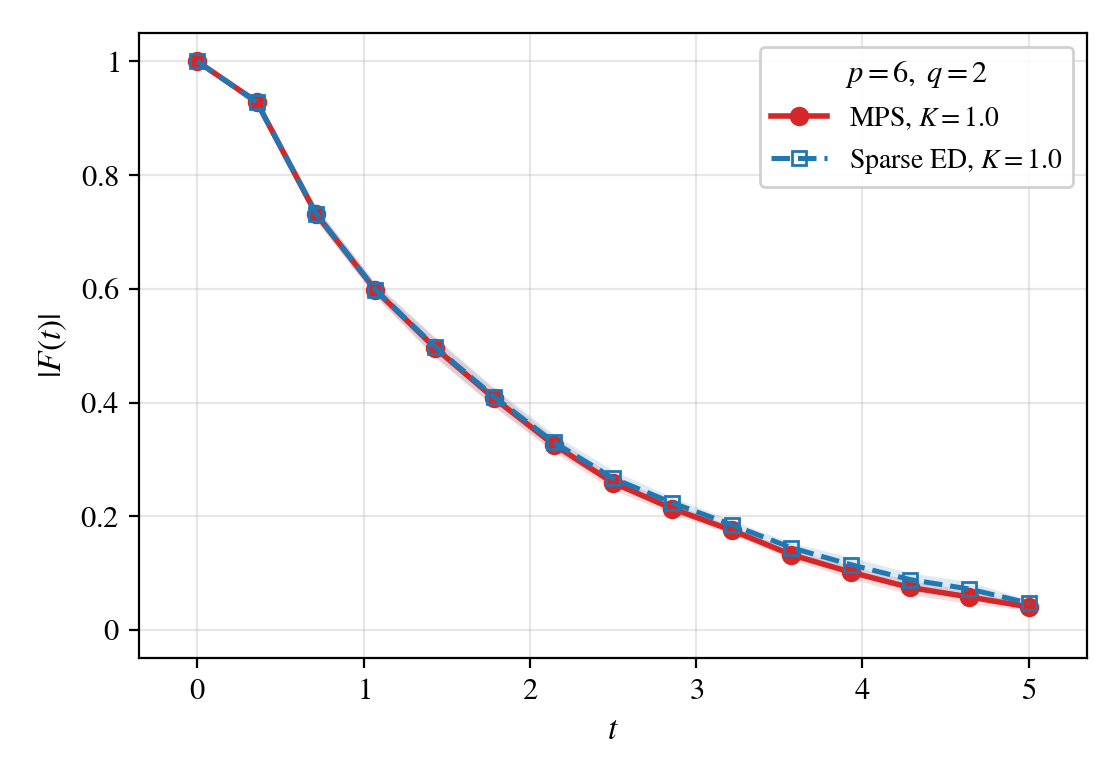}
\caption{OTOC magnitude $|F(t)|$ for $(p,q)=(6,2)$, $J=K=1$: MPS
($\chi=64$) versus ED.}
\label{fig:otoc_validation}
\end{subfigure}\hfill
\begin{subfigure}[t]{0.49\textwidth}
\centering
\includegraphics[width=\linewidth]{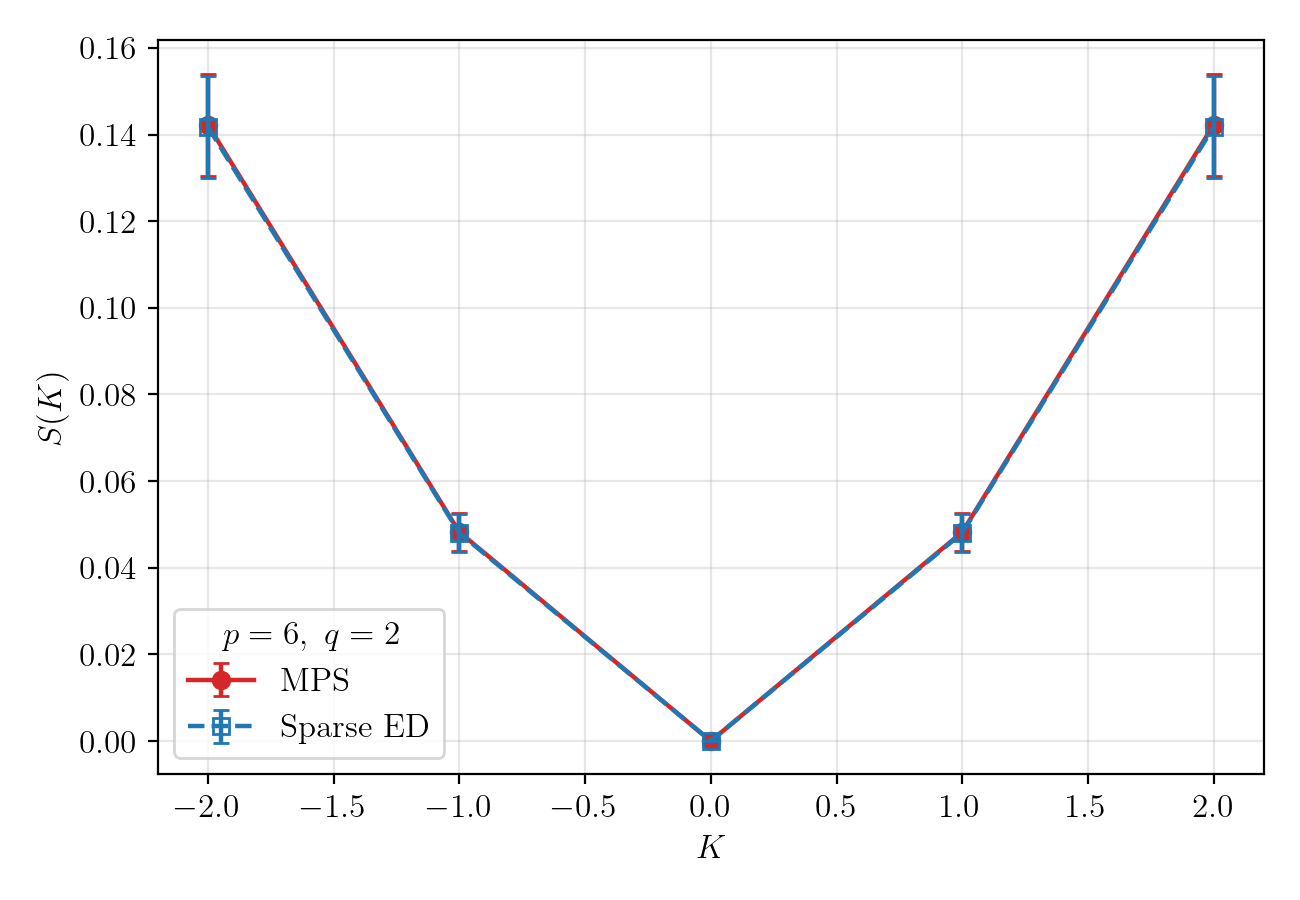}
\caption{Ground-state entanglement entropy $S_0$ versus $K$, for
$(p,q)=(6,2)$, $J=1$: MPS ($\chi=64$) versus ED.}
\label{fig:ee_validation}
\end{subfigure}
\caption{Validation against exact diagonalisation at $(p,q)=(6,2)$.
}
\label{fig:validation}
\end{figure*}

Here, we compare the MPS with ED. The ED is
constructed from the same sparse Pauli-string representation of $H$.
As
$\chi_{\mathrm{exact}}=64$ at this size, the MPS with $\chi\ge64$ is free of
truncation. This comparison tests the Trotterised per-dot propagators and
the convergence of the imaginary-time evolution.
 Further, Fig.~\ref{fig:ee_validation} shows that the MPS and ED values of $S_0$ agree
within the width of the lines at every $K$, with $S_0=0$ at $K=0$, where the
ground state is a product over dots. Similarly, Fig.~\ref{fig:otoc_validation}
shows that the MPS reproduces the OTOC of ED at $K=1$, to within about $0.02$
up to $t=5$. Thus, both the ground-state preparation and the real-time evolution are validated at a size where ED is an exact
reference.

\subsection{Bond dimension}
\label{sec:bond}

\begin{figure}[t]
\centering
\includegraphics[width=\linewidth]{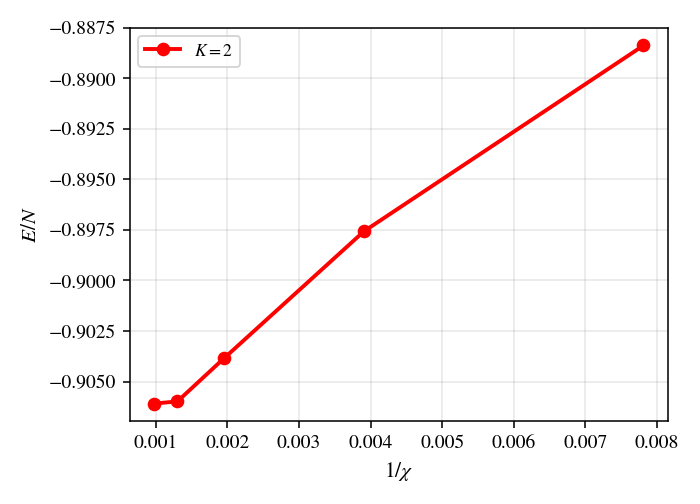}
\caption{Ground-state energy per site versus $1/\chi$ at
$(p,q)=(8,5)$, $J=1$ and $K=2$, for a single disorder realisation.}
\label{fig:N40}
\end{figure}

The resource estimates assume that the bond dimension is
$\chi\approx2^{\lfloor p/2\rfloor}\chi_{\mathrm{link}}$, with a link bond
dimension $\chi_{\mathrm{link}}$ set by $K$ (Sec.~\ref{sec:cost_mps}). We now test
this. Having validated the algorithm, we now turn to larger systems. We first
consider two geometries of comparable size but different structure. The first
is $(p,q)=(6,3)$, which has two links, both of which touch the middle dot. The
second is $(p,q)=(8,2)$, which has one link and larger dots, with
$\binom84=70$ rather than $\binom64=15$ quadruplets per dot. In both cases, we
use $J=1$ and $K\in\{-2,-1,0,1,2\}$. Further, we consider $(p,q)=(8,5)$, which
has the same $p$ as $(p,q)=(8,2)$. At this size, ED is beyond practical reach
(Sec.~\ref{sec:cost_methods}), and we consider $K=0$ and $K=2$,
where $K=2$ is the coupling with the largest
inter-dot entanglement among those studied (Sec.~\ref{sec:ee}).

We first fix the bond dimension for $(p,q)=(6,3)$ and $(8,2)$. For a single realisation, the energy
converges at $\chi_{\mathrm{conv}}=256$ for $(p,q)=(6,3)$ and
$\chi_{\mathrm{conv}}=64$ for $(p,q)=(8,2)$, i.e., at
$\chi_{\mathrm{exact}}/2$ and $\chi_{\mathrm{exact}}/4$
(Appendix~\ref{app:convergence}). Disorder-averaged energy (Table~\ref{tab:energy_avg}), entanglement entropy (Fig.~\ref{fig:ee_scaling_p8q2}) 
(Sec.~\ref{sec:otoc}), are computed at $\chi=256$ for $(6,3)$, $128$ for $(8,2)$ and $768$ for $(8,5)$,  and also OTOC at $\chi=256$ for $(6,3)$, $128$ for $(8,2)$, i.e., at or above $\chi_{\mathrm{conv}}$.

We now compare $(p,q)=(8,5)$ with $(p,q)=(8,2)$, at the same $p$ and $K$. Figure~\ref{fig:N40}
shows the energy per site at $(p,q)=(8,5)$ for $\chi=128$ to $1024$. The
energy changes by $1\times10^{-4}$ per site between $\chi=768$ and $1024$,
such that $\chi_{\mathrm{conv}}=768$, which is far below
$\chi_{\mathrm{exact}}=2^{20}$. However, this is larger than
$\chi_{\mathrm{conv}}=64$ at $(p,q)=(8,2)$, i.e., for this realisation, the
required bond dimension grows with $q$. By Eq.~\eqref{eq:rankbound},
$\chi_{\mathrm{conv}}=768$ requires $\chi_{\mathrm{link}}\ge48$, as the
inner bonds of a middle dot are at most $2^{\lfloor p/2\rfloor}=16$ times its
link bonds. In contrast, both dots of the $(8,2)$ chain are end dots,
whose inner bonds are also bounded by the number of qubits to the chain end,
such that $\chi_{\mathrm{conv}}=64$ is consistent with
$\chi_{\mathrm{link}}$ between $16$ and $64$. Thus, the growth of
$\chi_{\mathrm{conv}}$ is consistent with the factor
$2^{\lfloor p/2\rfloor}$ of the middle dots, and this realisation does not
show whether $\chi_{\mathrm{link}}$ itself grows with $q$.

\subsection{Ground-state energy}
\label{sec:energy}

\begin{table}[!htbp]
\caption{Disorder-averaged ground-state energy per site $\bar E/N$ (mean
$\pm$ standard error of the mean) versus $K$, for
$(p,q)=(6,3)$ and $(p,q)=(8,2)$, $J=1$.
}
\label{tab:energy_avg}
\begin{ruledtabular}
\begin{tabular}{ccc}
$K$ & $(p,q)=(6,3)$ & $(p,q)=(8,2)$ \\
\hline
$-2$ & $-0.8365\pm0.0076$ & $-0.9085\pm0.0056$ \\
$-1$ & $-0.8077\pm0.0077$ & $-0.8939\pm0.0056$ \\
$0$  & $-0.7982\pm0.0078$ & $-0.8889\pm0.0057$ \\
$1$  & $-0.8071\pm0.0081$ & $-0.8940\pm0.0056$ \\
$2$  & $-0.8359\pm0.0083$ & $-0.9090\pm0.0055$ \\
\end{tabular}
\end{ruledtabular}
\end{table}

The resource estimates treat the dots as the hard part, and the links
as weak perturbations. We now check this in the energy.
Table~\ref{tab:energy_avg} lists the disorder-averaged energy per site. We
find that the energy decreases monotonically with $|K|$ for both geometries,
and the values at $+K$ and $-K$ agree within the statistical error, as
required by the symmetry discussed in Sec.~\ref{sec:symmetries}. However,
the fluctuations of $E/N$ from one realisation to another are of the same
order as its dependence on $K$. As these fluctuations come mainly from the
dots, which are the same for every $K$ within a realisation, we quantify the
effect of the links through \emph{paired} differences within each
realisation, and define the energy gain per link as
\begin{equation}
\delta_{\mathrm{link}}(K)=\frac{N\,[E(K)-E(0)]}{q-1},
\label{eq:dlink}
\end{equation}
averaged over realisations. At weak coupling, we expect
$\delta_{\mathrm{link}}\simeq-cK^2$, as the first-order term
$K\langle Z_{(x+1)p-1}\rangle\langle Z_{(x+1)p}\rangle$ vanishes for even $p$
(Sec.~\ref{sec:symmetries}).
From the means over realisations, we obtain
$\delta_{\mathrm{link}}\approx-0.08$ at $|K|=1$ and $-0.31$ to $-0.35$ at
$|K|=2$, i.e., the two geometries give the same energy gain per link to
within about $10\%$. Further, $\delta_{\mathrm{link}}(2)/\delta_{\mathrm{link}}(1)$
is $4.0$--$4.2$ for $(6,3)$ and $3.9$ for $(8,2)$, consistent with
$\delta_{\mathrm{link}}\propto K^2$. At $(p,q)=(8,5)$, i.e., $N=40$, we obtain $\bar E/N=-0.8813\pm0.0025$ at $K=0$ and $-0.9020\pm0.0022$ at $K=2$, close to the values at $(8,2)$ (Table~\ref{tab:energy_avg}). Thus, the energy per site barely changes when
three dots are added.

\subsection{Entanglement across the link}
\label{sec:ee}

\begin{figure}[t]
\centering
\includegraphics[width=\linewidth]{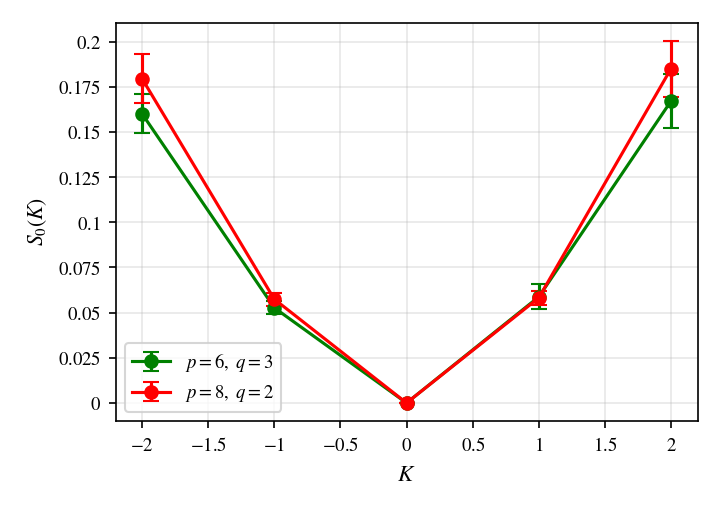}
\caption{Disorder-averaged ground-state entanglement entropy $S_0$ across the
dot-0 $|$ rest cut versus $K$, for $(p,q)=(6,3)$ and $(p,q)=(8,2)$, $J=1$.
Error bars are standard errors of the mean.}
\label{fig:ee_scaling_p8q2}
\end{figure}

The entanglement budget, and with it the classical cost, rests on a
small inter-dot entanglement set by $K$. 
Figure~\ref{fig:ee_scaling_p8q2} shows the
disorder-averaged entanglement entropy $S_0$ across the first inter-dot cut.
We find that $S_0$ vanishes at $K=0$, grows with $|K|$, and is even in $K$
within the statistical error, as required by Sec.~\ref{sec:symmetries}.
The two geometries give closely comparable entropies at every $K$,
i.e., $S_0\approx{}$ $0.05$--$0.06$ at $|K|=1$ and
$S_0\approx0.16$--$0.18$  at $|K|=2$, even though the number of
quadruplets per dot differs by a factor of almost five. At $(p,q)=(8,5)$, we obtain $\bar S_0=0$ at $K=0$, as expected for a product over dots, and $\bar S_0=0.200\pm0.014$ at $K=2$, close to the value at
$(8,2)$. Thus, the inter-dot entanglement is a comparatively local property, set mainly by $K$, which is the premise of the resource estimates. Indeed, even at $|K|=2$, $S_0$ remains small compared with the intra-dot scale $\lfloor p/2\rfloor\ln2$, i.e., $2.1$ for $p=6$ and $2.8$ for $p=8$.

\begin{figure}[t]
\centering
\includegraphics[width=\linewidth]{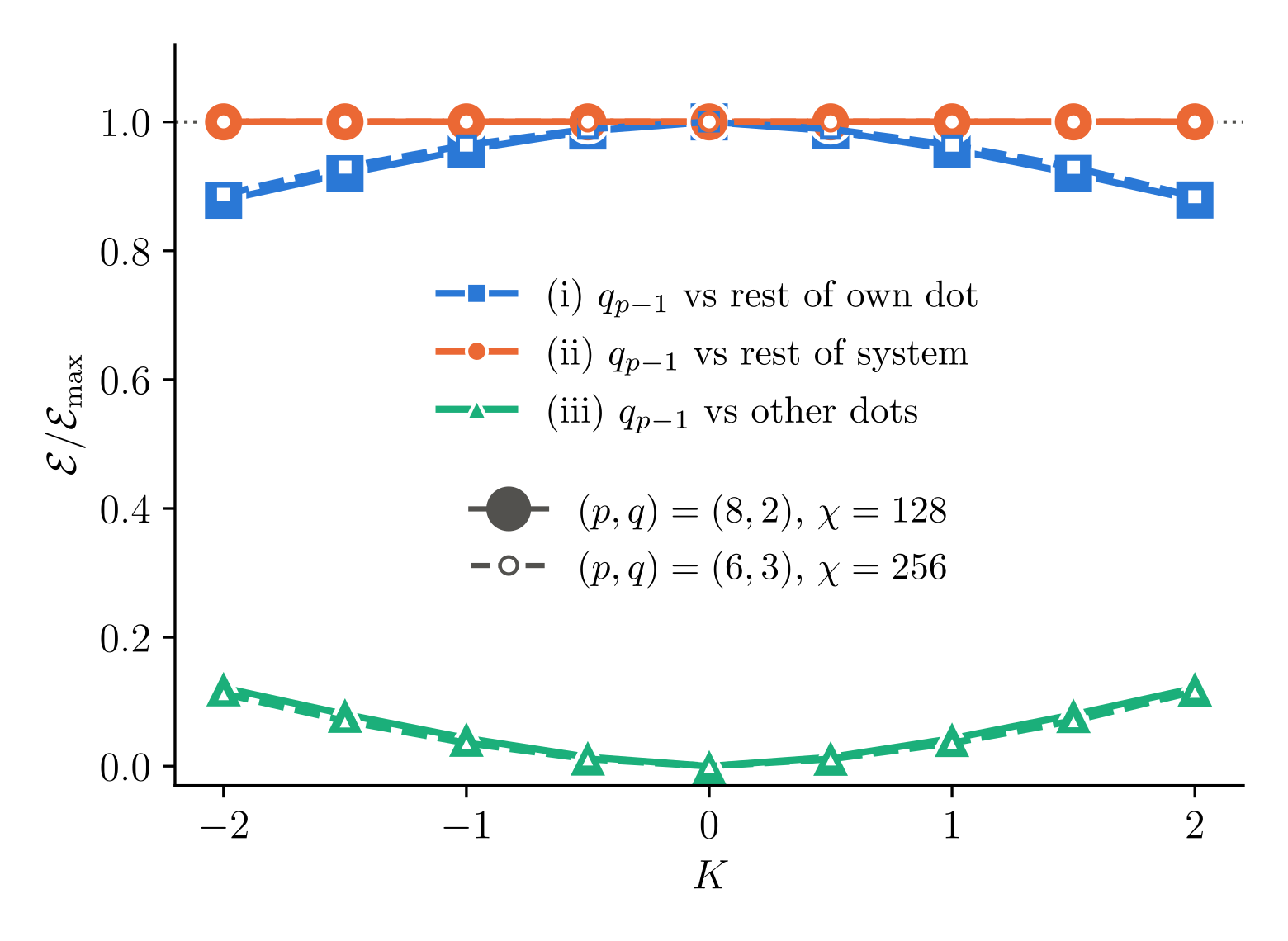}
\caption{Entanglement of the boundary qubit $q_{p-1}$ versus $K$, for
$(p,q)=(6,3)$ and $(8,2)$, $J=1$: $I(q_{p-1}\!:\!D_0')/{(2\ln2)}$ with the rest of
its own dot, $I(q_{p-1}\!:\!R)/{(2\ln2)}$ with the other dots, and $S(q_{p-1})/\ln2$ with
the rest of the system. For a single disorder realisation at
$\tau=213$ (Appendix~\ref{app:qubitent}).}
\label{fig:qubitent}
\end{figure}

We now turn from the entanglement of a whole dot to that of a single qubit at
the link. We consider the boundary qubit $q_{p-1}$ of dot $0$, and ask how its
entanglement is shared between the rest of its own dot, $D_0'$, and the other
dots, $R$. We quantify the two shares by the mutual informations
$I(q_{p-1}\!:\!D_0')$ and $I(q_{p-1}\!:\!R)$, which add up to $2S(q_{p-1})$
[Eq.~\eqref{eq:Isum}]. Importantly, for even $p$,
$S(q_{p-1})=\ln2$ at every $K$, as the link preserves the dot parities and the
symmetry $A$ (Sec.~\ref{sec:symmetries} and
Appendix~\ref{app:qubitent}). Thus, the boundary qubit stays maximally
entangled, and the link can only redistribute its entanglement.
Figure~\ref{fig:qubitent} shows the two shares for both geometries for a
single realisation. At $K=0$, $q_{p-1}$ is entangled only with its own dot. As
$|K|$ increases, the share with the other dots grows at the
expense of that with its own dot, to $0.11$--$0.12$ at $|K|=2$, and the
two geometries agree to within $0.01$ at every $K$. Further,
$q_{p-1}$ carries only a part of the correlation between dot $0$ and the other
dots, $I(D_0\!:\!R)=2S_0$, namely $36$--$41\%$ for $p=6$ and $28$--$31\%$
for $p=8$, nearly independent of $K$. Thus, although the link acts only on
$q_{p-1}$, the dot spreads most of the inter-dot correlation over its other
qubits.

\subsection{Operator spreading across the links}
\label{sec:otoc}

 The OTOC
defined in Eq.~\eqref{eq:otoc} quantifies the spreading of an operator that is
initially localised on the last qubit of the chain as it propagates across the
links towards the first qubit. Figure~\ref{fig:otoc} in
Appendix~\ref{app:otocfit} shows the disorder-averaged quantity $|F(t)|$ in
the ground state for both geometries. At $K=0$, the dots are dynamically
decoupled and $|F(t)|=1$ exactly, as discussed in
Sec.~\ref{sec:observables}.

For $K\neq0$, $|F(t)|$ remains close to unity at short times before undergoing
a monotonic decay, reaching values below $0.05$ by $t=7$. To characterise this
decay quantitatively, we fit each curve to the empirical form described in
Appendix~\ref{app:otocfit}. At late times, this form reduces to a stretched
exponential,
\begin{equation}
|F(K,t)|\simeq
\exp\!\left[
-\left(\frac{t-t_0(K)}{\tau_F(K)}\right)^{\beta(K)}
\right],
\label{eq:late_time}
\end{equation}
where $t_0(K)$ denotes the onset time, $\tau_F(K)$ is the decay time, and
$\beta(K)$ is the shape exponent. Increasing the coupling from $|K|=1$ to
$|K|=2$ reduces $\tau_F$ by a factor of $5.6$ for $(p,q)=(6,3)$ and by
approximately $4.0$ for $(p,q)=(8,2)$. Over the same range, $\beta$ decreases
from approximately $1.4$--$1.5$ to $0.6$--$0.7$, indicating a change from an
almost Gaussian decay to a sub-exponential one.

These results show that, although a single link contributes only weakly to the
ground-state entanglement discussed in Sec.~\ref{sec:ee}, it can be crossed
by an operator on a time scale that decreases rapidly with increasing $|K|$.
The curves for the $(p,q)=(6,3)$ geometry, in which the operator must traverse
two links, remain close to unity for a longer time, corresponding to the larger
onset times $t_0$ reported in Table~\ref{tab:fit_params}. However, because the
two geometries also have different values of $p$, this delay cannot be
attributed solely to the number of links.
\section{Discussion}
\label{sec:discussion}

Locating a regime in which quantum hardware could outperform classical simulation needs a model that stays tractable on both sides, and a parameter that moves it towards the classically hard regime. We have designed such a model, the SYK-dot chain, in which $q$ dense bosonic SYK dots of $p$ qubits are joined by single $ZZ$ bonds of strength $K$. The physical content of this construction is the entanglement budget of Eq.~\eqref{eq:entbound}: for any pure state, the entanglement across any cut is bounded by $\lfloor p/2\rfloor\ln 2$ plus the largest inter-dot entanglement, with no explicit dependence on $N=pq$. In the ground state, the inter-dot entanglement is static and set by $K$, and we find it to be small. Thus, the dot size $p$ sets the hardness on both sides, while $K$ controls the inter-dot entanglement.

\paragraph{Where a quantum advantage could appear.}
Because the dots act on disjoint qubits and the links join only neighbours, the two-qubit depth of one Trotter step on hardware is set by $p$ alone [Eq.~\eqref{eq:depth}]. After transpilation onto a heavy-hex device, the depth stays close to that of a single dot up to $N=30$ for $p=6$ and $N=40$ for $p=8$, although for $p=8$ it is two to two and a half times the all-to-all value of Eq.~(\ref{eq:depth}). Dense SYK reaches $5.0\times10^{5}$ at $N=20$, and at the sizes studied the sparse model of Ref.~\cite{Gautam25} with $\kappa=2$ is still shallower than the chain with $p=8$ (Fig.~\ref{fig:circuitcost}). On the classical side, the cost of the block MPS grows linearly in $q$ at a fixed bond dimension [Eq.~\eqref{eq:tmps}], and in our timings is lower than that of a conventional gate-by-gate MPS by a factor of $40$ to $60$ at $p=6$ and $8$. However, its bond dimension can grow as $2^{\lfloor p/2\rfloor}\chi_{\mathrm{link}}$, and an MPS with one site per dot is estimated to be cheaper for larger dots. Taking an MPS with one site per dot as the classical reference (estimated from the operation count, not run), and a budget of $10^{18}$ operations and 1~TB RAM, a quantum advantage for the ground state could appear only for dots of about 15 to 21 qubits or more at $q=5$. This assumes $\chi_{\mathrm{link}}$ between 16 and 256, which we have estimated only for $p\le 8$. Further, the estimate counts the hardware cost of a single Trotter step, and thus assumes that the number of steps needed to prepare the ground state, e.g., adiabatically, does not grow exponentially with $p$. A single Trotter step requires two-qubit error rates of order $10^{-6}$ or below (Sec.~\ref{sec:cost_map}). In contrast, NISQ hardware with error rates of $10^{-3}$ to $10^{-4}$ can run a single step only for $p\le5$ to $8$, where classical simulation is cheap. We do not claim an advantage at the sizes studied here; instead, we locate where one could appear, and how far it lies beyond NISQ hardware.

\paragraph{Tests of the assumptions.}
Our MPS results test the assumptions behind this analysis. The imaginary-time MPS agrees with ED at $(p,q)=(6,2)$. At $(p,q)=(8,5)$, i.e., $N=40$ qubits, where ED is out of reach, the ground-state energy of a single realisation at $K=2$ converges at bond dimension $\chi=768$, far below the exact value $2^{20}$ (Fig.~\ref{fig:N40}). This increase from $\chi=64$ at $(8,2)$ is consistent with Eq.~\eqref{eq:rankbound} for the middle dots, which appear only for $q\ge3$, and does not show whether $\chi_{\mathrm{link}}$ grows with $q$. The disorder-averaged energy per site and $S_0$ at $(8,5)$ are close to those at $(8,2)$, at both $K=0$ and $K=2$. Testing the $q$ dependence would need several values of $q$ and more disorder realisations. Even at $|K|=2$, the ground-state entropy across a link, $S_0$, is small compared with the intra-dot scale $\lfloor p/2\rfloor\ln 2$ ($2.1$ for $p=6$, $2.8$ for $p=8$), and it is nearly the same for $(6,3)$ and $(8,2)$, although the number of quadruplets per dot differs by almost a factor of five (Fig.~\ref{fig:ee_scaling_p8q2}). This is consistent with a single $ZZ$ bond generating only a limited amount of entanglement between two dots whose internal states are already fixed by their own strong interactions. Further, the link only redistributes the entanglement of the boundary qubit, and the dot spreads most of the inter-dot correlation over its other qubits (Fig.~\ref{fig:qubitent}), which suggests that the strong interactions within a dot spread the effect of a single bond over the whole dot. For even $p$, the first-order term $K\langle Z_{(x+1)p-1}\rangle\langle Z_{(x+1)p}\rangle$ vanishes, so the energy gain per link, defined in Eq.~\eqref{eq:dlink}, is expected to start at second order in $K$.

\paragraph{Variational preparation under noise.}

At $(p,q)=(5,2)$, a variational circuit with the dot-plus-link layout,
trained classically, comes within $3.6\%$ of the exact energy with $88$ CNOT
gates. depolarising gate noise, its fidelity decays approximately as
$e^{-\epsilon N_{\mathrm{2q}}}$, as assumed in Sec.~\ref{sec:cost_map}.
After routing onto a linear chain of qubits, its energy stays within $10\%$
of ED only for $\epsilon\lesssim3\times10^{-4}$, i.e., below the error rates
of current devices (Sec.~\ref{sec:noisy}).

\paragraph{Operator spreading.}
Real-time dynamics leave this controlled regime. The OTOC of Eq.~\eqref{eq:otoc} decays on a time scale that shrinks rapidly with $|K|$: in the empirical fit of Eq.~\eqref{eq:model}, the decay time $\tau_F$ drops by a factor of about $4$ to $6$ between $|K|=1$ and $2$ (Fig.~\ref{fig:otoc} and Table~\ref{tab:fit_params}). However, this speed-up cannot continue at strong coupling. Every SYK term that acts on a boundary qubit flips it in the $Z$ basis, and thus changes the energy of the link by $2|K|$. For $|K|\gg J$, this cost is much larger than the interaction strength within the dots, such that these terms are suppressed and the boundary qubits are effectively frozen. An operator can then no longer cross the link easily, and we expect the decay of the OTOC to slow down again at large $|K|$. We have studied only $|K|\le 2J$, and have not tested this.

\paragraph{Limitations.}
First, we have not characterised the chaos of the individual dots, e.g., through level statistics within each parity sector. Second, the noisy simulations at $(5,2)$ use a single realisation at $K=2$, and include only gate errors. Third, at $(8,5)$, the convergence in $\chi$ is tested for a single realisation, and the disorder averages use fewer realisations than at $(8,2)$, such that the dependence on $N$ rests on one comparison with $(8,2)$. Fourth, our imaginary-time energies are upper bounds, with an error of order $10^{-5}$ per site at $(p,q)=(5,2)$, and the real-time OTOCs carry a Trotter error that we have not quantified. Fifth, preparing the ground state on hardware
beyond small variational circuits, e.g., adiabatically, has a cost set by the
gap of the dots, which we have not studied. Finally, our results concern ground states and the spreading of operators across the links. In long-time dynamics, the inter-dot entanglement is no longer fixed by
$K$, but grows with time (Sec.~\ref{sec:budget}), and so does the bound of
Eq.~\eqref{eq:entbound}.

\paragraph{Outlook.}
On the hardware side, real-time dynamics from a product state, e.g.,
quenches or infinite-temperature OTOCs, need no ground-state preparation, and
the depth per Trotter step of Sec.~\ref{sec:cost_step} applies directly.
We believe that they are natural targets for hardware runs, with our ground-state results as a reference. Further, testing the dot-plus-link ansatz at larger
$q$ would show whether its accuracy, like its depth, is independent of $q$. On the physical side, adding bonds between neighbouring dots, or coupling the dots on more general graphs, would increase the inter-dot entanglement. We expect this to move the model continuously towards the volume-law regime, and thus to locate the boundary of classical tractability in a controlled way. On the numerical side, a second-order Trotter decomposition would lower the $O(\Delta\tau)$ floor of the energy, and a larger $\chi$ would give access to later times in the OTOC.
\begin{acknowledgments}

This work was supported by the Anusandhan National Research Foundation (ANRF), India, under grant ANRF/ECRG/2025/003752/PMS; by MICIU/AEI/10.13039/501100011033 (project PID2022-136231NB-I00) and the European Regional Development Fund (FEDER, EU); and by the BasQ and IKUR Strategies through the collaboration agreement between the Ikerbasque Foundation and the Donostia International Physics Center (DIPC), on behalf of the Department of Education of the Basque Government. The authors acknowledge the C-DAC HPC facilities under the National Supercomputing Mission (NSM) for computing resources, and the DIPC Computer Center for technical and human support.

\end{acknowledgments}

\section*{Data availability}
The data are available from the authors upon reasonable request.

\appendix

\section{Derivation of the entanglement budget}
\label{app:budget}

In this appendix, we derive Eqs.~\eqref{eq:entbound} and~\eqref{eq:rankbound}. Since the ground state is pure, $S(A)=S(\bar A)$ for any region $A$ and its complement $\bar A$, so a cut may be compared with
a nearby boundary by transferring qubits on either side. If $A$ is the region on one side of a cut and $C$ is a block of $m$ qubits adjacent to it, then subadditivity and the Araki--Lieb inequality give
\[
|S(A)-S(C)|\leq S(AC)\leq S(A)+S(C),
\]
and hence
\[
|S(AC)-S(A)|\leq S(C)\leq m\ln 2 .
\]
That is, moving a block of $m$ qubits from one side of a bipartition to the
other changes the entropy by at most $m\ln 2$.

The cut after the first $r$ qubits of dot $x$ ($0\le r\le p$) differs from
the boundary between dots $x-1$ and $x$ by $r$ qubits, and from the boundary
between dots $x$ and $x+1$ by $p-r$ qubits. Therefore,
\begin{align}
S_{\mathrm{cut}}
  &\leq \min\bigl\{\, S_{x-1} + r\ln 2,\;\;
                       S_x + (p-r)\ln 2 \,\bigr\}
  \label{eq:cut-min}\\[2pt]
  &\leq \max_x S_x + \min\{r,\,p-r\}\ln 2 .
  \label{eq:cut-max}
\end{align}
In Eq.~\eqref{eq:cut-max}, both $S_{x-1}$ and $S_x$ have been bounded by
$\max_x S_x$, which can then be taken outside the minimum. Finally,
$\min\{r,\,p-r\}\leq\lfloor p/2\rfloor$ for every integer $0\le r\le p$, which
gives Eq.~\eqref{eq:entbound}.

The same argument applies to Schmidt ranks, as moving a block of $m$ qubits
across a cut changes the Schmidt rank by at most a factor $2^m$. Thus, the
bond dimension at the cut obeys
\begin{equation}
\chi_{x,r} \le \min\{2^r\chi_{x-1},\;2^{p-r}\chi_x\}
\le 2^{\lfloor p/2\rfloor}\max\{\chi_{x-1},\chi_x\},
\end{equation}
which is Eq.~\eqref{eq:rankbound}.

\section{Convergence in bond dimension and imaginary time}
\label{app:convergence}

\begin{figure*}[t]
\centering
\begin{subfigure}[t]{0.49\textwidth}
\centering
\includegraphics[width=\linewidth]{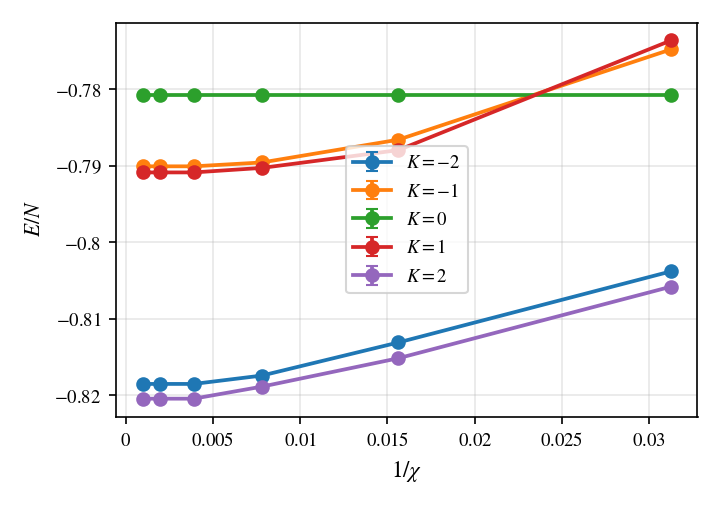}
\caption{$(p,q)=(6,3)$, $\chi_{\mathrm{exact}}=512$.}
\label{fig:energy_scaling}
\end{subfigure}\hfill
\begin{subfigure}[t]{0.49\textwidth}
\centering
\includegraphics[width=\linewidth]{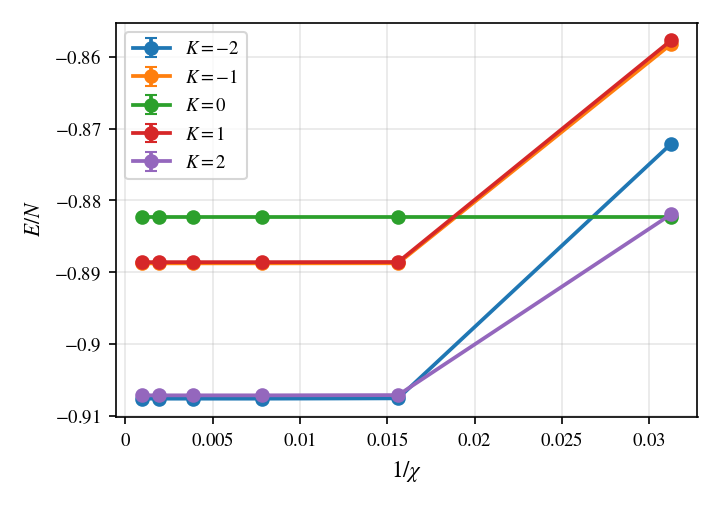}
\caption{$(p,q)=(8,2)$, $\chi_{\mathrm{exact}}=256$.}
\label{fig:energy_scaling_p8q2}
\end{subfigure}
\caption{Ground-state energy per site versus $1/\chi$ at
$K\in\{-2,-1,0,1,2\}$, $J=1$, for a single realisation.
}
\label{fig:chi_convergence}
\end{figure*}

In this appendix, we discuss the convergence of our ground states. We first
consider the bond dimension. Figure~\ref{fig:chi_convergence} shows the
ground-state energy per site at $(p,q)=(6,3)$ and $(8,2)$ as a function of
$1/\chi$ for a single disorder realisation, for $\chi$ from $32$ to $1024$. This gives the values quoted
in Sec.~\ref{sec:bond}. For both geometries, the energy at $K=0$ is
converged already at $\chi=32$, as the state is a product over dots and only
the intra-dot entanglement must be represented.

We now turn to the imaginary-time evolution. Besides $\chi$, two parameters control the state reached. First, the
$O(\Delta\tau)$ Trotter error of Sec.~\ref{sec:imagtime} sets a floor on
the energy error, which cannot be removed by increasing $\tau$ or $\chi$.
Second, $\tau$ must be long compared with $1/\Delta$, where $\Delta$ is the
gap to the lowest level outside the ground-state manifold. At $K=0$, this is
the gap of a single dot. However, for odd $p$, the dot doublets are split only
at first order in $K$, such that $\Delta\propto|K|$ is small at weak coupling.

We have checked the resulting error at $(p,q)=(5,2)$, where the MPS is free of
truncation. There, the MPS energy lies above the ED energy by
$1\times10^{-5}$ per site at $K=2$ ($E_{\mathrm{MPS}}=-0.66799$ and
$E_{\mathrm{ED}}=-0.66800$ for a single realisation). Thus, our ground-state energies are upper bounds, with an error of
order $10^{-5}$ per site at this size.

Finally, we note that the curves at $K=+2$ and $K=-2$ in
Fig.~\ref{fig:energy_scaling} differ at every $\chi$, including
$\chi\ge\chi_{\mathrm{exact}}$, where the MPS is exact. Thus, this difference
does not arise from truncation, but from the breaking of the $K\to-K$ symmetry
at fixed disorder (Sec.~\ref{sec:symmetries}), and its size is further
affected by the finite imaginary time discussed above.
\subsubsection{Bond-dimension convergence of the energy at $p=6$}
\label{sec:chiconv}

We now test how the converged bond dimension depends on $q$ at fixed $p=6$.
For $q=2$, $3$ and $4$, we compute the ground-state energy per site at
increasing $\chi$ for a single disorder realisation at $K=2$, and take
$\chi_{\mathrm{conv}}$ as the smallest tested $\chi$ at which the energy agrees
with the converged value{(see above)}. As the $\chi$ grid is coarse,
$\chi_{\mathrm{conv}}$ is an upper bound on the true threshold.
Table~\ref{tab:chiconv} lists the results.

\begin{table}[t]
\caption{Bond dimension at which the ground-state energy is converged at
$p=6$, for a single realisation. $\chi_{\mathrm{conv}}$ is the smallest tested
$\chi$ that meets the tolerance, so it is an upper bound on the true
threshold.}
\label{tab:chiconv}
\begin{ruledtabular}
\begin{tabular}{ccccc}
$(p,q)$ & $N$ & $\chi_{\mathrm{exact}}$ & $\chi_{\mathrm{conv}}$ & $\chi_{\mathrm{conv}}/\chi_{\mathrm{exact}}$ \\
\hline
$(6,2)$ & $12$ & $2^{6}$  & $\le 64$  & $\le 1$ \\
$(6,3)$ & $18$ & $2^{9}$  & $\le 256$ & $\le 1/2$ \\
$(6,4)$ & $24$ & $2^{12}$ & $\le 512$ & $\le 1/8$ \\
\end{tabular}
\end{ruledtabular}
\end{table}
At $p=6$ and $K=2$, the entanglement across the first link is nearly
independent of $q$: a single realisation gives $S_0=0.2255$ at $q=3$ and
$S_0=0.2267$ at $q=4$, i.e., it changes by about $0.5\%$ when a dot is added.
This is the behaviour assumed in the resource estimates (Sec.~\ref{sec:cost_map}), in which the inter-dot entanglement is set by $K$
and not by $N$. At the same time, the converged bond dimension stays well
below $\chi_{\mathrm{exact}}$ and its ratio to $\chi_{\mathrm{exact}}$
decreases with $q$ (Table~\ref{tab:chiconv}), so the MPS remains far cheaper
than exact methods. The upper bound on $\chi_{\mathrm{conv}}$ increases with $q$ in this
range. From $q=2$ to $3$, this can reflect the middle dot, for which
Eq.~\eqref{eq:rankbound} allows an additional factor $2^{\lfloor p/2\rfloor}$.
From $q=3$ to $4$, both chains have middle dots, but only at $q=4$ does a link
have two dots on each side, such that its $\chi_{\mathrm{link}}$ is no longer
capped by the end dot, $\chi_{\mathrm{link}}\le2^p$. A single realisation on a
coarse grid cannot separate this growth of $\chi_{\mathrm{link}}$ from the grid
resolution.

\section{Entanglement of the boundary qubit}
\label{app:qubitent}
In this appendix, we describe how we compute the entanglement of the boundary
qubit $q_{p-1}$ of dot $0$, and how it is shared between
its own dot and the other dots. We denote the other $p-1$ qubits of dot $0$ by
$D_0'$, and the qubits of all other dots by $R$. As the MPS is a pure state,
the mutual informations of ${q_{p-1}}$ with $D_0'$ and with $R$ are
\begin{align}
I(q_{p-1}\!:\!D_0') &= S(q_{p-1})+S(D_0')-S(D_0), \label{eq:Iown}\\
I(q_{p-1}\!:\!R) &= S(q_{p-1})+S(D_0)-S(D_0'), \label{eq:Iout}
\end{align}
where $D_0=q_{p-1}\cup D_0'$ and $S$ is the von Neumann entropy. Thus,
\begin{equation}
I(q_{p-1}\!:\!D_0')+I(q_{p-1}\!:\!R)=2S(q_{p-1}),
\label{eq:Isum}
\end{equation}
i.e., the entanglement of ${q_{p-1}}$ is shared between its own dot and the other
dots.

As $D_0$ and $D_0'$ are the first $p$ and $p-1$ sites of the MPS, we obtain
$S(D_0)$ and $S(D_0')$ from the Schmidt values $\lambda_i$ at the bonds
$(p-1,p)$ and $(p-2,p-1)$, $S=-\sum_i\lambda_i^2\ln\lambda_i^2$. For each bond,
we first bring a copy of the MPS into mixed canonical form with its
orthogonality centre on that bond, as the stored singular values are Schmidt
values only in this form. Further, $S(q_{p-1})$ follows from the Bloch vector
$r_\mu=\langle\sigma^\mu_{p-1}\rangle$, as $\rho_{q_{p-1}}$ has the eigenvalues
$(1\pm|\bm r|)/2$. At $(p,q)=(5,2)$ and $(6,2)$, all three entropies agree with
those of the full state vector to within $10^{-13}$.

For even $p$, $S(q_{p-1})=\ln2$ in the ground state at every $K$, as $r_x=r_y=0$ in
a parity eigenstate and $r_z=0$ by the symmetry $A$
(Sec.~\ref{sec:symmetries}). However, $|{+}\rangle^{\otimes N}$ overlaps
with every parity sector, and the Trotter factors do not mix them. Thus, the
evolved MPS reaches a parity eigenstate only for long $\tau$. We
therefore use a longer evolution for these runs, up to $\tau=213$. With it, $|1-S(q_{p-1})/\ln2|<4\times10^{-6}$ at every $K$ for $(p,q)=(6,3)$ and $(8,2)$.

\section{Fit of the OTOC decay}
\label{app:otocfit}

\begin{figure*}[t]
\centering
\begin{subfigure}[t]{0.49\textwidth}
\centering
\includegraphics[width=\linewidth]{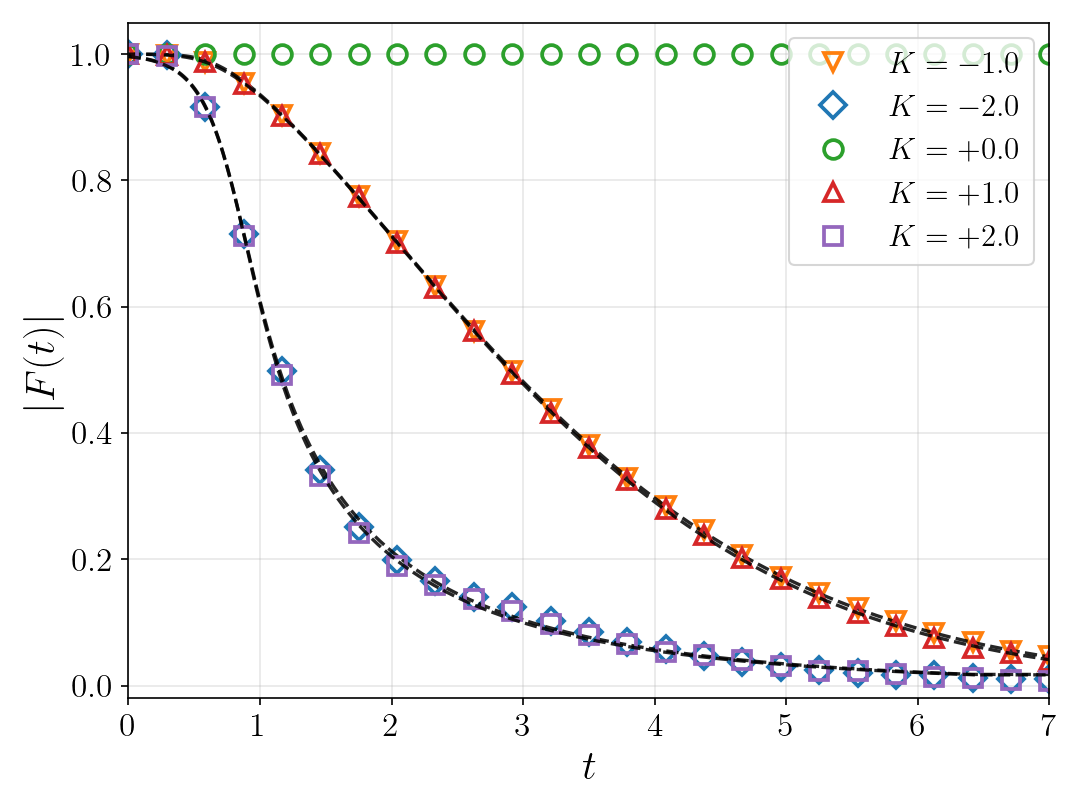}
\caption{$(p,q)=(6,3)$, $\chi=256$.}
\label{fig:otoc_chi256}
\end{subfigure}\hfill
\begin{subfigure}[t]{0.49\textwidth}
\centering
\includegraphics[width=\linewidth]{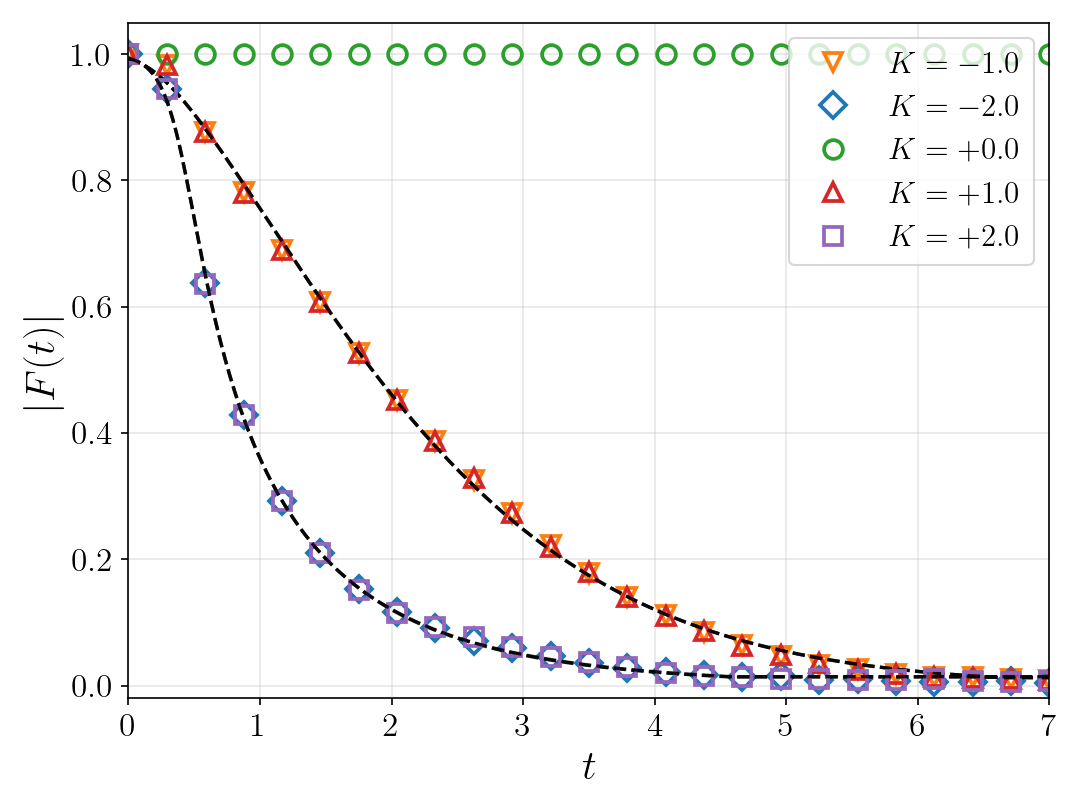}
\caption{$(p,q)=(8,2)$, $\chi=128$.}
\label{fig:otoc_p8q2}
\end{subfigure}
\caption{OTOC magnitude $|F(t)|$, Eq.~\eqref{eq:otoc}, in the MPS ground state
for $K\in\{-2,-1,0,1,2\}$, $J=1$, and
averaged over disorder realisations at the converged bond dimension
(Sec.~\ref{sec:bond}). Markers show the data, and dashed lines show
the empirical fits of Eq.~\eqref{eq:model} for $K\neq0$, with the parameters
listed in Table~\ref{tab:fit_params}.
}
\label{fig:otoc}
\end{figure*}

\begin{table*}[!htbp]
\caption{Parameters of Eq.~\eqref{eq:model}, with $t_0,\tau_F,\beta,\nu$ functions of $|K|$ only, fitted jointly to the $K=\pm1,\pm2$ traces for each geometry. As $\pm K$ share their parameters, the RMS residual is listed for each sign.
$^{*}$ $\nu$ is only weakly constrained at $(8,2)$, $|K|=1$: the fit is nearly
flat in $\nu$ there, so this value should not be over-interpreted.}
\label{tab:fit_params}
\begin{ruledtabular}
\begin{tabular}{ccccccc}
$(p,q)$ & $|K|$ & $t_0$ & $\tau_F$ & $\beta$ & $\nu$ & RMS residual ($-K$ / $+K$) \\
\hline
$(6,3)$ &$1$ & $0.490$ & $3.072$ & $1.522$ & $5.54$ & $2.8\times10^{-3}$ / $2.9\times10^{-3}$ \\
$(6,3)$ &$2$ & $0.846$ & $0.549$ & $0.603$ & $9.38$ & $7.5\times10^{-3}$ / $7.5\times10^{-3}$ \\
\hline
$(8,2)$ &$1$ & $0.035$ & $2.347$ & $1.427$ & $65.4^{*}$ & $8.6\times10^{-3}$ / $8.7\times10^{-3}$ \\
$(8,2)$ &$2$ & $0.394$ & $0.584$ & $0.727$ & $15.8$ & $2.9\times10^{-3}$ / $3.6\times10^{-3}$ \\
\end{tabular}
\end{ruledtabular}
\end{table*}

Figure~\ref{fig:otoc} shows the disorder-averaged OTOCs of
Sec.~\ref{sec:otoc}. To characterise the decay of $|F(K,t)|$ as a function of both the coupling $K$ and
the time $t$, we fit every curve with $K\neq 0$ to the empirical form
\begin{align}
|F(K,t)| &= \exp\!\left[-\left(\frac{s_{\nu(K)}\big(t-t_0(K)\big)}{\tau_F(K)}\right)^{\beta(K)}\right], \label{eq:model}\\[4pt]
s_{\nu}(x) &= \frac{1}{\nu}\ln\!\big(1+e^{\nu x}\big), \label{eq:softplus}
\end{align}
where the softplus function $s_{\nu}(x)$ tends to $\max(x,0)$ as
$\nu\to\infty$ and removes the kink at $t=t_0$. Thus, a nearly flat window
($t<t_0$, where $|F|\approx1$) is followed by a decay with rate $\tau_F^{-1}$ and shape $\beta$, where $\beta=1$ and $2$ give exponential and Gaussian decay, respectively. All four parameters depend on $|K|$, since the fitted values change systematically with $K$. Eq.~\eqref{eq:model} is a \emph{phenomenological parametrisation} of the data,
and is not derived from an underlying microscopic (e.g., Schwinger--Dyson)
equation for this model. For $t-t_0\gg1/\nu$, $s_\nu(t-t_0)\to t-t_0$, such that Eq.~\eqref{eq:model} reduces to the stretched exponential of
Eq.~\eqref{eq:late_time}.

Table~\ref{tab:fit_params} lists the fitted parameters. The fit reproduces
every curve with a root-mean-square (RMS) residual of at most
$8.7\times10^{-3}$, and the residuals at $+K$ and $-K$ are close, as
expected after disorder averaging (Sec.~\ref{sec:symmetries}). The
changes of $\tau_F$ and $\beta$ with $|K|$ are discussed in
Sec.~\ref{sec:otoc}. The fitted $t_0$ increases with
$|K|$. However, as $t_0$, $\tau_F$ and $\beta$ are correlated in the fit, $t_0$
alone is not the onset of scrambling, and the curves at $|K|=2$ decay faster
overall.

\bibliography{refs}

\end{document}